\documentclass[12pt,preprint]{aastex}

\usepackage{natbib}

\shorttitle{Star Formation at High Galactic Latitudes}

\shortauthors{H.-T. Lee and J. Lim}

\begin{document}

\title{ON THE FORMATION OF PERSEUS~OB1 AT HIGH GALACTIC LATITUDES}

\author{Hsu-Tai Lee\altaffilmark{1} \email{htlee@asiaa.sinica.edu.tw}}

\author{Jeremy Lim\altaffilmark{1} \email{jlim@asiaa.sinica.edu.tw}}

\altaffiltext{1}{Institute of Astronomy and Astrophysics, Academia Sinica, P.O. Box 23-141, Taipei 106, Taiwan, R.O.C.}

\begin{abstract}
The Per~OB1 association, which contains the remarkable double cluster $h$ and $\chi$~Per, is unusual in not having a giant molecular cloud in its vicinity.  We show from {\it Hipparcos} data that the luminous members of this association exhibits a bulk motion away from the galactic plane, such that their average velocity increases with height above the galactic plane.  We find HAeBe and T~Tauri stars towards probable remnant molecular clouds associated with Per~OB1.  These star-forming regions lie well beyond the location of the luminous member stars at heights of 280--400~pc above the galactic plane, far higher than that previously found for embedded clusters.  We argue that the observed motion of the luminous member stars is most naturally explained if many formed from molecular gas pushed and accelerated outwards by an expanding superbubble driven presumably by stellar winds and perhaps also supernova explosions.  A large shell of atomic hydrogen gas and dust that lies just beyond the remnant molecular clouds, believed to be driven by just such a superbubble, may comprise the swept-up remains of the parental giant molecular cloud from which this association formed.  In support of this picture, we find a week trend for the younger O star members to lie at higher galactic latitudes than the older supergiant members.  The star-forming regions located at even larger heights above the galactic plane presumably correspond to more recent episodes of star formation at or near the periphery of this superbubble.

\end{abstract}


\section{INTRODUCTION}

O and B stars are not distributed at random in the sky, but instead are usually concentrated in loose co-moving groups that \citet{amb47} first referred to as OB associations.  OB associations are generally distributed along spiral arms, and confined in the galactic plane.  They have low stellar densities of order 0.1~M$_{\sun}$~pc$^{-3}$, and are easily disrupted by Galactic tidal forces \citep{bok34,amb47} so must be relatively young with ages $\lesssim 30$~Myr \citep[see, e.g., reviews by][]{dez99}.

Some of the best evidence that the formation of massive stars can induce the formation of other low- to high-mass stars have been found in OB associations.  On scales up to a few tens of parsecs, massive stars can ionize the surfaces of small molecular clouds (globules) in their vicinity \citep{sug91}; the pressure from such ionization fronts can then propagate through the globule to help overcome the magnetic, turbulent, and thermal pressure that supports it against collapse, thereby inducing star formation.  This process of radiation-driven implosion has been invoked to trigger star formation in many globules within OB associations \citep{rei83}, not least by one of us in the Orion region \citep{lee05,lee07a} and Lac OB1 \citep{lee07a}.  On the other hand, star-forming globules in OB associations may simply be dense cores that have spontaneously collapsed to form stars, and were then uncovered by advancing ionization fronts from nearby massive stars.  In the examples described in \citet{lee05} and \citet {lee07a}, classical T~Tauri stars (CTTSs)  are found only between OB stars and bright-rimmed (i.e., ionized surface of) molecular clouds, not within these clouds; furthermore, those CTTSs closer to the bright-rimmed clouds are progressively younger, arguing for a genuine sequence in star formation.

On larger scales of tens of parsecs or more, expanding superbubbles driven by ionizing radiation and stellar winds from massive stars, together perhaps with supernova explosions, have been invoked to both sweep up and compress gas, and subsequently induce star formation \citep[see, e.g., review by][]{pre07}.  Such large-scale triggerings of star formation may be required to explain co-moving OB subgroups with either identical ages or arranged in an apparent age sequence, with at least some subgroups having an age spread that is much smaller than their stellar crossing time.  One of the best such examples is the Scorpius-Centaurus (ScoCen) OB association \citep[see, e.g., review by][]{pre07}, which is the closest OB association to the Sun.   It comprises the three co-moving subgroups Upper Scorpius (USco) with an age of $\sim$5~Myr, Upper Centaurus-Lupus (UCL) with an age of $\sim$17~Myr, and Lower Centaurus-Crux (LCC) with an age of $\sim$16~Myr distributed from approximately east to west in galactic longitude over a spatial extent of $\sim$130~pc \citep{deg89,mam02}.  The age dispersion in the USco subgroup (the youngest OB subgroup) is no more than $\sim$1--2~Myr, compared with a stellar crossing time of $\sim$20~Myr \citep{pre02}.  \citet{deg89} and \citet{pre99} proposed that subsequent supernova explosions in LCC induced star formation in the molecular cloud that formed USco.

The abovementioned picture of triggered star formation is based on an updated variation of the collect-and-collapse model of \citet{elm77}, where a propagating disturbance first collects gas into dense clouds and subsequently induces star formation.  Here we present our study of the Perseus~OB1 (Per~OB1) association, which as shall show has properties that can be most naturally explained by the cloud-shuffling model of triggered star formation as first proposed by \citet{elm79}.  Cloud shuffling takes place on the periphery of an OB association where a superbubble pushes and accelerates a giant molecular cloud outwards, triggering star formation at the dense cloud interface \citep{elm79,elm92}.  Until now, Cepheus~OB3 (Cep~OB3) has been regarded as one of the best examples where star formation has been triggered by cloud shuffling \citep{sar79,elm92}.  Cep~OB3 contains two OB subgroups, the younger ($\sim$4~Myr) of which is closer to and has a radial velocity resembling the Cep~B molecular cloud, and the older ($\sim$ 8~Myr) further away and has a radial velocity resembling the rest of the cloud.  \citet{sar79} and \citet{elm92} proposed that ionizing radiation and stellar winds from the older OB subgroup pushed and accelerated the Cep~B cloud away, as well as induced star formation in this cloud to form the younger subgroup.

Our paper is organized as follows.  In \S2 we collect together previous studies that compile the proposed members of Per~OB1, and describe our study of the proper motion of those members included in the {\it Hipparcos} catalog; we also describe our search for low to intermediate-mass pre-main-sequence stars towards Per~OB1, and our identification of possible remnant molecular clouds in this association.  In \S3 we interpret and discuss our results in the context of current star formation models.  In \S4 we briefly summarize the most important results of our study and their chief implications.

\section{DATA AND RESULTS}

\subsection{Candidate Luminous Members}

The Per~OB1 association, located in the region of the Perseus spiral arm, is commonly (but not without argument) thought to be centered about the $h$ and $\chi$~Per double cluster \citep[see discussion in][]{sle02}.  Widely referred to as the double cluster, $h$ and $\chi$~Per are among the brightest, densest, and closest of open clusters containing moderately massive (main-sequence B, along with supergiant) stars.  Both clusters have been studied extensively over the last century \citep[see references in][]{sle02}.  The latest study by \citet{sle02} conclude that the two clusters have essentially identical ages of $13 \pm 1$~Myr, distances of $2.34 \pm 0.05$~kpc, comparable initial masses (integrating between 1--120~M$_{\sun}$) of 3700~M$_{\sun}$ for $h$~Per and 2800~M$_{\sun}$ for $\chi$~Per, and a Salpeter IMF.

Although containing more than ten O stars, Per~OB1 is not associated with any known \ion{H}{2} regions.  Furthermore, unlike virtually all other OB associations studied in molecular gas, Per OB1 is not associated with any known giant molecular cloud.  \citet{sle02} quote a failed attempt to detect molecular gas in unpublished $^{13}$CO observations by L. A. Hillenbrand \& J. M. Carpenter using the FCRAO.  In addition, they find that the reddening to the $h$ and $\chi$ Per double cluster is indistinguishable and produced entirely by line-of-sight extinction to the Perseus spiral arm, with the intracluster reddening essentially zero.

\citet{hum78} collected from the literature all the then known early and late-type supergiant stars, O stars, and both main-sequence as well as giant B0--B1 stars that were proposed to be members of this association.  She recomputed their photometric distances using the then latest available luminosity calibration for O and B stars, as well as for late-type supergiants.  A total of 105 members were identified, all of which also have comparable radial velocities.  All the member stars lie within the range in galactic longitude of $132\degr \leq l \leq 136\degr$ and galactic latitude of $-5\degr \leq b \leq -2\degr.5$.

The most recent compilation from the literature of all proposed early-type luminous members of Per OB1 was made by \citet{gar92}.  By studying where the early-B (B0--B2) main-sequence stars lie on the color-magnitude diagram (i.e., cluster fitting method), \citet{gar92} derived a distance modulus for related members of $\sim$11.8, which corresponds to a distance of $\sim$2.3~kpc.  Next to trigonometric parallaxes, this is the most secure method for identifying cluster members.  \citet{gar92} also identified possible members from among the early-type giant to supergiant stars based on the then best available color-magnitude relations.  In this way, they identified a total of 149 members in Per OB1, including $\sim$$70\%$ of the stars listed in \citet{hum78} as members of this association.  Note that \citet{gar92} only considered early-type (O--A) stars, whereas the cataloged members of \citet{hum78} also include late-type (supergiant) stars.  The Per OB1 members identified by \citet{gar92} span a significantly larger range in galactic longitude of $131\degr \lesssim l \lesssim 139\degr$ as well as galactic latitude of $-6\degr.5 \lesssim b \lesssim 0\degr$.

In Figure~1, we plot the proposed members of Per~OB1 as cataloged by \citet{hum78} and \citet{gar92} in Galactic coordinates.  [We exclude HD~237019, whose position has been incorrectly entered in Table~2a of \citet{gar92}.  This star is a member of Cas OB6 \citep{hum78}.]  A clustering of B stars (green circles) can be seen towards two closely separated positions, corresponding to the $h$ and $\chi$ Per double cluster (indicated by large black circles).  No clear clustering is seen for the O stars (blue circles), nor the supergiant stars (red circles).  The relative deficiency in cluster members at low galactic latitudes ($b \gtrsim -2\degr.5$) appears to be real; as can be seen in Figure~1 of \citet{gar92} for Per OB1, many of the early-type stars at $b > -2\degr.5$ were found to be either field stars or members of other associations.  On the other hand, it is not clear to us whether the identification of luminous early-type stars is as complete at low galactic heights (given the larger extinction and greater confusion) as at larger heights.

From a search of the literature since the work by \citet{gar92}, we have found three more luminous stars toward Per OB1 that have a comparable distance modulus as inferred by \citet{gar92} for this association.  These are the WC6 star HD~17638 \citep[distance 1.91~kpc,][]{van01}, the O9.5III(n) star HD~15137 \citep[distance 2.65~kpc,][]{van88}, and the ON8V star HD~14633 \citep[distance 2.15~kpc,][]{van88}.  The locations of these three stars are indicated in Figure 2.  The radial velocities of HD~15137 and HD~14633 \citep{boy05} are closely comparable with the range observed for the Per OB1 association.  These two stars, especially HD~14633, lie at a significantly higher galactic latitude than those cataloged by either \citet{hum78} or \citet{gar92}.

\subsection{Hipparcos Proper Motions}

The Hipparcos astrometry mission \citep{kov84} permits a study of the proper motion of stars in (or towards) Per OB1.  Note that the stars in this association are too far away for Hipparcos to obtain a measurement of their trigonometric parallax, which is predicted to be $\sim$0.4~mas compared with measurement uncertainties of $\sim$1~mas \citep{per97}.

Sixty-one out of a total of 180 stars cataloged by \citet{hum78} and \citet{gar92} as members of Per OB1, as well as the three new additions mentioned in $\S2.1$, are contained in the Hipparcos catalog.  The positions of these stars, converted to galactic coordinates, are listed in Table~\ref{table:members} together with their Hipparcos catalog (Hip) number, proper motion in Galactic longitude and latitude along with their associated standard errors, star name in the Henry-Draper (HD) or Bonner Durchmusterung (BD) catalog, spectral type, and whether these stars are listed as members by \citet{hum78} or \citet{gar92} or both.  Most of these stars are either early- or late-type giants and supergiants.  

In Figure~2, we plot the proper motion of the stars listed in Table~\ref{table:members} in Galactic coordinates.  The blue, green, and red arrows are O, B, and supergiant stars respectively.  The arrows are plotted with lengths proportional to the magnitude of their proper motions: note, however, that the uncertainty in proper motion changes from star to star, and the uncertainty in galactic longitude can be quite different to that in galactic latitude.  We also plot the proper motion of the $h$ and $\chi$~Per clusters (black arrows) as determined by \citet{kha05} based on space mission and ground-based observations.  h~Per has a proper motion in ($l$, $b$) of ($-0.16 \pm0.37$~mas yr$^{-1}$, $-1.01 \pm0.13$~mas yr$^{-1}$), and $\chi$~Per a proper motion of ($-0.71 \pm0.47$~mas yr$^{-1}$, $-0.51 \pm0.17$~mas yr$^{-1}$).  For comparison, we also plot the proper motions of all the other early-type (B0--B9) stars (yellow arrows) in the Hipparcos catalog within the field encompassed by Figure~2.  These stars are referred to hereafter as "field" stars, although some may be members of other clusters or associations, and yet others perhaps previously unidentified members of Per OB1 as we explain below.  The white arrow oriented at a position angle of $67^{\circ}$ (measured anticlockwise from Galactic north) indicates the direction towards the solar apex; i.e., the direction (as projected in the sky) towards which the Sun is moving relative to the local standard of rest.

As can be seen in Figure~2, some of the field stars (yellow arrows) have very large proper motions that in many cases lie at position angles of roughly $110\degr$.  These are relatively nearby stars (as confirmed from an examination of their trigonometric parallaxes) that, because of our motion relative to the local standard of rest, appear to be moving in a direction opposite to that of the Sun (white arrow in Fig.~2).  The proposed members of the Per~OB1 association, being much more distant, do now show any such appreciable motion.

To study the actual proper motion of the proposed Per OB1 members relative to the galaxy at their (presumed) distance (2.3~kpc) and direction, we subtract our motion relative to the local standard of rest and galactic rotation at a distance of 2.3~kpc in the direction of the individual members.  The correction for $h$~Per is 0.27~mas yr$^{-1}$ in $l$ and $0.75$~mas yr$^{-1}$ in $b$, giving an actual proper motion of ($0.11 \pm0.37$~mas yr$^{-1}$,  $-0.26 \pm0.13$~mas yr$^{-1}$); for $\chi$ Per, the correction is $0.24$~mas yr$^{-1}$ in $l$ and $0.75$~mas yr$^{-1}$ in $b$, giving an actual proper motion of ($-0.47 \pm0.47$~mas yr$^{-1}$, $0.24 \pm0.17$~mas yr$^{-1}$).  Thus, the double cluster does not exhibit significant motion relative to the galaxy rotation and galactic plane at its location.

We tabulate the average measured proper motion of the remaining luminous member stars in Galactic latitude strips of width $1\degr$ from latitudes of $0\degr$ to $-8\degr$ in Table~\ref{table:pm}.  (Note that the measured proper motion of the individual stars is in general of low statistical significance).  The average proper motion in longitude is not significantly (i.e., $\lesssim 3 \sigma$) different from zero except for the two individual stars at the two highest latitudes (HD~15137 at $b \approx -7\degr.6$ and HD~14633 at $b \approx -18\degr.2$).  The average proper motion in latitude is significantly (at least 4$\sigma$) different from zero at latitudes $b \lesssim -2\degr$, including the two individual stars at the two highest latitudes.  To better visualize the results, in Figure~3 we plot the proper motion of the individual members in galactic latitude, as well as their proper motion averaged over the different latitude strips.  Apart from the abovmentioned motion away from the Galactic plane at $b \lesssim -2\degr$, there is also a clear trend towards larger motions away from the Galactic plane at higher latitudes, with the star at the highest latitude (HD~14633) exhibiting the largest proper motion away from the Galactic plane.  Note that the members of Per OB1 cataloged by \citet{hum78} have in most cases radial velocities in the range $-40 {\rm \ km \ s^{-1}}$ to $-50 {\rm \ km \ s^{-1}}$, the same range in radial velocities as the Perseus spiral arm.  We therefore conclude that the bulk motion of the stars in Per~OB1 is predominantly away from the galactic plane, with velocities up to a few tens of km s$^{-1}$ for most of the stars with measured proper motions.

A visual inspection of Figure~2 reveals that many of the "field" stars at latitudes $-3\degr.5 \gtrsim b \gtrsim -4\degr.5$ (yellow arrows) spanning approximately the same longitude range as the proposed association members also exhibit a preferential motion away from the galactic plane.  The average proper motion of the stars in this latitude strip and in the longitude range $138\degr \lesssim l \lesssim 128\degr$ is ($0.26\pm0.12$~mas/yr, $-1.87\pm0.12$~mas/yr).  Their average proper motion in Galactic latitude is comparable with the values tabulated in Table~\ref{table:members} for the proposed member stars in this latitude strip, and hence these stars are likely genuine members of Per~OB1.  Most of these stars have spectral types in the range B5--B9, and would therefore not have been included in the catalogs of either \citet{hum78} or \citet{gar92}.

\subsection{Candidate Low to Intermediate Mass PMS Stars}

\subsubsection{Color Selection}

To study whether there has been any recent (within the last $\sim$1~Myr or so, compared with an age for the $h$ and $\chi$ Per double cluster for $\sim$ 13~Myr) formation of low- to intermediate-mass stars in Per~OB1, we have searched the Two Micron All-Sky Survey (2MASS) point source catalog \citep{cut03} for pre-main-sequence (PMS) star candidates.  Our selection of such candidate stars was based on the empirical criteria developed by \citet{lee07a} for low-mass CTTSs and intermediate-mass Herbig Ae/Be (HAeBe) stars, but with improved criteria here for better selecting the latter stars.  With this method, we first select stars showing near-infrared (near-IR) excess thereby preferentially picking out PMS stars.  Because HAeBe stars generally exhibit a stronger near-IR excess than CTTS, we then separated these two stellar classes based on their 2MASS colors as described below.  Our initial list comprised all stars in the 2MASS catalog at $121\degr < l < 147\degr$ and $-22\degr < b < 2\degr$ \citep[i.e., spanning a much wider area in the sky than the members of Per~OB1 as identified by][]{gar92} that exhibit a near-IR excess.

In Figure~4, we plot the (J-H) versus (H-K) color diagram for  selecting PMS star candidates.  CTTS candidates are those defined to lie between the two parallel lines $(J-H)-1.7(H-K)+0.0976=0$ and $(J-H)-1.7(H-K)+0.450=0$, and above the dereddened CTTS locus $(J-H)-0.493(H-K)-0.439=0$ \citep{mey97}.  HAeBe candidates are defined as those lying between the two parallel lines $(J-H)-1.7(H-K)+0.450=0$ and $(J-H)-1.7(H-K)+1.400=0$, and above the line $(J-H)=0.2$.  The line $(J-H)=0.2$ helps discriminates classical Be stars from HAeBe stars.  A full explanation for the adopted criteria will be presented elsewhere \citep{lee07b}.  In this way, we found a total of 399 candidate CTTS and 264 candidate HAeBe stars.

In Figure~5, we plot the spatial distribution of the candidate PMS stars (red for candidate CTTS, green for candidate HAeBe stars, and blue for early-type members of Per~OB1).  At low galactic latitudes of $b \lesssim -5\degr$, apart from three clusters associated with the known HII regions W3, W4, and W5 of the Cas OB6 association, the candidate PMS candidates (which are presumably confused with cool evolved stars) are roughly equally distributed in galactic longitude.

\subsubsection{Candidate Remnant Molecular Clouds}

In Figure~5, we also plot (in grayscale) the $IRAS$ 100-$\micron$ map (dominated by emission from interstellar dust) of the Per OB1 region.  The map in the right-hand panel has been stretched to accentuate features at high galactic latitudes ($b \lesssim -5\degr$) coincident with the candidate PMS stars selected in the manner described above.  As can be seen, around the region occupied by the Per OB1 association, a number of the candidate PMS stars at galactic latitudes as low as about $-10\degr$ appear to be close to or coincident with dust clouds.  An examination of optical images from the 2nd Digitized Sky Survey (DSS2) as shown in Figure~6 reveals bright-rimmed clouds, comet-shaped clouds, and reflection clouds coincident with these dust clouds.

The abovementioned clouds seen in the DSS2 images resemble remnant molecular clouds (RMCs; Ogura et al. 1998).  RMCs are small molecular clouds that can be commonly found in the vicinity of OB associations.  Their outlines are shaped by ionization fronts from massive stars, and hence they often have cometary morphologies with heads pointing towards the ionizing source.  There is at least one such previously known example in Per OB1, referred to as Red Nebulous Object~6 \citep[RNO6;][]{coh80}.  \citet{bac02} found two molecular clouds, each having a mass of $\sim$$200 {\rm \ M_\sun}$, around RNO6.  They showed that these clouds have cometary morphologies with their heads pointing towards the $h$ and $\chi$~Per clusters.  They also confirmed that RNO6 hosts an embedded IR cluster that includes a HBe star responsible for illuminating RNO6, and suggested that the recent star formation in this RMC was triggered by radiatively driven implosion.  As can be seen in Figure~6, we find another candidate HAeBe apparently associated with RNO6.  (The known HBe star in RNO6 does not exhibit a near-IR excess in 2MASS.)

Based on Figure~5, we have identified four other candidate RMCs towards Per OB1 that appear to be closely associated with candidate PMS stars.  (We emphasize that our selection method does not provide a complete census of RMCs in this region.)  They are henceforth referred to as RMC1--RMC4 with locations as indicated in Figure~5 and tabulated in Table~\ref{table:rmc}.  A search of the catalog of bright nebulae compiled by \citet{lyn65} reveals that RMC1, RMC2, and RMC3 are associated with LBN~638, LBN~640, and LBN~666 respectively.  All four candidate RMCs have cometary morphologies as shown in Figure~5 and 6 with heads pointing towards the galactic plane, rather than to any particular O stars in Per OB1.

\subsubsection{Spectroscopic Observations}

To determine if there are other sites of recent star formation in Per OB1 apart from RNO6, we performed spectroscopic observations of the candidate PMS stars close to or coincident with the abovementioned candidate RMCs (including RNO6).  The location of these stars are plotted in Figure~6, and tabulated in Table~\ref{table:spectra}.  We also observed four stars apparently associated with reflection nebulae located in the vicinity of, and likely associated with, the RMCs.  These stars do not exhibit detectable near-IR excess in 2MASS, and hence are not included in our list of candidate PMS stars.  The locations of these stars also are shown in Figure~7, and tabulated in Table~\ref{table:spectra}.  These stars will prove useful in obtaining a distant estimate to the RMCs.

We obtained spectra of both the candidate PMS stars and stars associated with reflection nebulae at the Beijing Astronomical Observatory (BAO) in 2005 and 2006.  Low-dispersion spectra with a dispersion of 200~\AA~mm$^{-1}$ or 4.8~\AA~pixel$^{-1}$ were taken with a 2.16-m optical telescope on 2005 December~6--8.  We used an OMR (Optomechanics Research, Inc.) spectrograph together with a Tektronix 1024$\times$1024 CCD detector covering the wavelength range 4000-9000~\AA.  On 2006 November~11--13, we used the same spectrograph but now deployed with a new Princeton Instrument SPEC10 400$\times$1300B CCD detector camera covering the wavelength range 3500-8500~\AA.

We processed the spectroscopic data in the standard manner using the NOAO/IRAF package.  After correcting for bias and flat-fielding, we used the IRAF program KPNOSLIT to extract and calibrate the wavelength and intensity of each spectrum.  The results for the individual stars are presented in Table~\ref{table:spectra}.  CTTS and HAeBe stars are identified by the presence of H$\alpha$ emission, and separated by the presence of other Balmer lines seen in absorption in HAeBe stars.  We separate intermediate-mass stars into HAeBe, B, and A stars based on the spectral classification scheme of \citet{jac84}.  We do not, however, attempt to determine the spectra types of CTTS because their spectra exhibit heavy veiling.

Of the nine candidate PMS stars selected by 2MASS colors and observed, we confirmed that four are HAeBe stars and five are possibly T Tauri stars.  Detection of the Li line, impossible in our low-dispersion spectra, is required to confirm the nature of the CTTS candidates.  We find PMS stars in the vicinity of or coincident with all four RMCs.  As there are no known foreground (or background) star-forming regions in their directions, and given their large heights above the galactic plane (see more detailed discussion in $\S2.3.4$), their association with the RMCs is likely genuine.  The RMCs are therefore likely to be regions of recent or ongoing star formation.  Of the four stars (stars 5, 6, 8, and 10) associated with reflection nebulae studied, three were found to be B main-sequence stars (stars 5, 6, and 10) and one an A main-sequence star (star 8).

\subsubsection{Distance to Candidate Star-Forming Sites}

We now use the B or A stars associated with reflection nebulae to determine a photometric distance to these stars and hence to two of the candidate RMCs.  Specifically, we will focus on 2MASS J01494099$+$5341341 (B7V; star 5 in Table~\ref{table:spectra}), 2MASS J01500456+5354007 (B3V; star 6) and 2MASS J01501546$+$5341350 (A2V; star 8) apparently associated with RMC2, and 2MASS J02220484$+$5038130 (B8V; star 10) with RMC3.

To derive the absolute magnitudes in the J-band of stars with the abovementioned spectral classes, we selected eleven B3V, three B7V, forty-one B8V, and 128 A2V stars with good parallax ($\pi$) measurements, [relative error in parallax of $\sigma_{\pi} \pi^{-1} < 10\%$ \citep{bro97}, goodness-of-fit parameter $<$ 3, and percentage of rejected data $< 10\%$ \citep{oud99}], and good 2MASS photometric qualities.  By selecting only stars with small relative errors in parallax, we hope to minimize the bias in determining absolute magnitudes and hence distances using this method.  As pointed out by \citet{lut73}, the inclusion of stars with large relative errors in parallax biases the final result to a lager parallax than the true parallax.

Based on measurements of their apparent magnitudes in the 2MASS catalog, we then derived median absolute magnitudes of $-$0.652, 0.131, 0.484, and 1.176 for the B3V, B7V, B8V, and A2V stars respectively in the J-band.  Based on their apparent magnitudes in the J-band in the 2MASS catalog, the corresponding distance modulus of 2MASS J01500456+5354007 (B3V) is 11.324, 2MASS J01494099$+$5341341 (B7V) is 11.831, and 2MASS J01501546$+$5341350 (A2V) is 12.457; i.e., a distance of  1840~pc, 2323~pc, and 3100~pc to RMC2.  Similarly, the distance modulus of 2MASS J02220484$+$5038130 (B8V) is 12.327; i.e., a distance of 2920~pc to RMC3.  All the estimated distance module agree with the range measured by \citet{hum78} to Per OB1 of magnitude 10.83 -- 12.97.  The inferred distances to RMC2 and RMC3 are therefore compatible with their location in Per~OB1.

As mentioned above, all four candidate RMCs appear to be associated with PMS stars.  If at the distance of the Per~OB1 association, then these star-forming regions are located at unprecedented heights above the galactic plane for known clusters embedded in molecular clouds.  In Figure~8, we plot the height above the galactic plan of seventy-six such embedded clusters compiled by \citet{lad03} within $\sim$ 2~kpc of the Sun, as well as the four RMCs studied here (assuming a distance to these RMCs of 2.3~kpc).  As can be seen, $\sim$$90\%$ of these clusters are located within $\sim$160~pc of the galactic plane.  Only two of the embedded clusters cataloged by \citet{lad03} are located beyond a height of 200~pc, NGC 281W and NGC 281E, both of which extend no one higher than 240~pc.  By comparison, all four RMCs studied here are situated more than 280~pc away from galactic plane, with RMC3 having the largest height of $\sim$390~pc.

\section{INTERPRETATION AND DISCUSSION}

As mentioned in $\S1$, Per~OB1 is unusual in that it is not associated with a giant molecular cloud.  Based on the results described in $\S2$, we now find that Per~OB1 is unusual in at least two other respects.  First, most of its members exhibit a bulk motion away from the galactic plane with those at larger heights moving faster (on average) away from the galactic plane.  Second, star formation has recently occurred (and may still be occurring) at unusually large heights of 280--400~pc above the galactic plane.  We now examine whether all of these features can be explained, and if so how, in the framework of current star formation models.

\subsection{Fate of Giant Molecular Cloud}

Star formation at the present epoch is thought to be relatively inefficient in converting molecular gas to stars, even more so in models that invoke spontaneous as opposed to triggered star formation.  To explain the lack of a giant molecular cloud associated with Per~OB1 without invoking an extraordinarily effective dispersement or destruction mechanism, one would have to propose that the star formation in this association has been more efficient than normal in converting molecular gas to stars.  This may, but is not necessarily required to, explain the unusually large stellar densities and masses of the $h$ and $\chi$~Per double cluster.  The high number density of stars and their resultant bipolar outflows, stellar winds, UV ionization, and/or supernova explosions would then be more effective also in sweeping away or destroying much of the remaining molecular gas.

In a search for shells and supershells in the Galaxy, \citet{hei79} identified a large shell of atomic hydrogen (\ion{H}{1}) gas in the direction towards Per~OB1.  \citet{cap00} show that this \ion{H}{1} shell occupies the same range in radial velocities as Per~OB1, and forms a quasi-circular structure of size $\sim$$350 \times 500$~pc centered at ($l, b$)=(134$\degr$.5, $-2\degr$.5).  The center of this shell therefore lies quite close to the locations of the $h$ and $\chi$~Per double cluster.   \citet{cap00} estimate a total mass of atomic gas in the shell of order $10^{5}$ M$_{\sun}$.  They infer an an upper limit of 10~km s$^{-1}$ for the expansion velocity of the \ion{H}{1} shell, and a kinetic energy for the shell of $9 \times 10^{49}$~ergs.  They also compute a minimum mechanical energy to blow the shell of $\sim 4.5 \times 10^{50}$ ergs, compared with an energy released through stellar winds by all the known member B stars of $\sim 1.3 \times 10^{51}$~ergs.  Thus, stellar winds from B stars alone in Per OB1 is sufficient to blow the observed \ion{H}{1} shell, let alone an additional energy from sources such as supernova explosions.  We note that the total kinetic energy of the member stars due to their motion away from the galactic plane ($\sim 2 \times 10^{49}$~ergs) adds insignificantly to the overall energetics.

\citet{cap00} show from the {\it IRAS} 100-$\mu$m image that emission from dust traces the inner border of the \ion{H}{1} shell in the portion at large heights above the galactic plane (any dust associated with the \ion{H}{1} shell close to the galactic plane is confused with galactic dust).  This dust feature can be seen as a nearly semi-circular arc in Figure~2 of \citet{cap00}, extending to a galactic latitude of about $-11^{\circ}$ at its largest height above the galactic plane.  We speculate that this dust shell may have formed from the swept-up giant molecular cloud.

We have checked the literature to see if the presumed hot bubble of gas responsible for sweeping up the abovementioned HI and dust shell is visible in X-rays.  Such X-ray emission is detectable in the ROSAT all sky survey (RASS) \citep{sno95a} from the Orion-Eridanus Superbubble \citep{bro95}, which is powered by Ori OB1 (which contains approximately the same number of O stars as Per OB~1).  This superbubble has approximately the same physical size as that enclosed by the HI shell associated with Per~OB1.  The kinematic energy of the Orion-Eridanus Superbubble, as derived from its associated expanding \ion{H}{1} shell, is about $3.7 \times 10^{51}$~ergs \citep{bro95}, which is between one and two orders of magnitude larger than that of the presumed Per OB1 Superbubble.  If the Orion-Eridanus Superbubble was located at the distance of Per OB1, the count rate range of 3/4~keV band would be about 5--15 counts s$^{-1}$.  By comparison, the background count rate towards the presumed Per OB1 Superbubble is about 110 counts s$^{-1}$.  Perhaps not surprisingly then, we did not find any significant X-ray emission from RASS in 1/4, 3/4, and 1.5~keV bands towards the presumed Per OB1 Superbubble.

We have searched the literature to find maps in molecular gas towards Per OB1.  \citet{dam01} has mapped in CO(1-0) the sky in the direction towards Per OB1 at an angular resolution of 8\farcm5.  Compact and isolated CO emission is detectable towards RMC1, RMC2, and RMC3, but not towards RMC4.  If these sources are genuinely associated with the abovementioned remnant molecular clouds, the corresponding molecular gas mass is in these clouds $\sim$ 8000~M$_{\sun}$.  No CO emission was detected towards RNO6, which has a molecular gas mass of $\sim$ 400~M$_{\sun}$, consistent with the upper limit in sensitivity of the \citet{dam01} map.   No CO emission was detected towards the dust/\ion{H}{1} shell either; assuming a CO linewidth of 10~km s$^{-1}$, the corresponding 3$\sigma$ upper limit in molecular gas along this shell is $\sim 10^{4}$ M$_{\sun}$.  If this estimate is accurate, the bulk of the molecular gas in this shell may therefore have been dissociated into atomic hydrogen gas.

\subsection{Acceleration away from Galactic Plane}

Much of the star formation in the Per OB1 association appears to have occurred in or within the vicinity of the $h$ and $\chi$~Per double cluster.  If so, then the observed increase (on average) in velocity with height (acceleration) away from the galactic plane may be a natural consequence of the dissolution of these clusters \citep{kro01,gey01,lad03}.  Over time, stars moving faster away from the double cluster would travel furthest from their initial locations, thus producing the observed linear increase in velocity with height above the galactic plane.

Although this model could explain the velocity pattern of the B main-sequence and supergiant stars, which can have ages up to $\sim$ 10~Myr of more, it cannot easily explain the same velocity pattern observed for the O stars.  The latter stars have ages of only a few millions years, much younger than the B stars in the $h$ and $\chi$ Per double cluster ($\sim$ 13~Myr), and could not have reached their present locations if traveling at their observed velocities from these clusters.  This model also predicts that the member stars should form an approximately spherical distribution moving away from the $h$ and $\chi$~Per double cluster, which is not seen (c.f. Fig.~2).

Given that the molecular gas from which the stars in Per~OB1 formed appears to have been effectively swept outwards by the luminous stars in this association ($\S3.1$), we now consider whether the observed velocity dependence with height could simply be a reflection of stars forming in this swept-up gas.  The situation envisaged is akin to that proposed in models of triggered star formation, in particular the scenario of cloud shuffling.  Cloud shuffling takes place on the periphery of an OB association where an over-pressured superbubble of accumulated stellar winds, supernova explosions, and/or ionized gas pushes and accelerates a giant molecular cloud outwards \citep{elm79,elm92}.  Such cloud shuffling is proposed to induce star formation, and also extend the lifetimes of molecular clouds by pushing them away from massive stars and their destructive effects.  In the present case, the dust shell at the inner boundary of the \ion{H}{1} shell presumably lies just beyond the expansion front of the superbubble associated with Per~OB1.

Assuming typical physical conditions in OB associations, \citet{elm79} suggests that cloud shuffling can accelerate molecular gas by about 1~km~s$^{-1}$~Myr$^{-1}$ (or equivalently 1~pc~Myr$^{-2}$).  Let us assume that the $h$ and $\chi$~Per double cluster was the original source of the superbubble (with luminous stars formed later from the swept-up molecular gas adding to the energy of this superbubble), consistent with its location close to the center of the \ion{H}{1} shell.  The age of this double cluster is estimated to be $\sim$13$\pm$1~Myr \citep{bra05,kel01,sle02}, allowing sufficient time for the molecular gas at the edge of the superbubble to be accelerated by over $10 {\rm \ km \ s^{-1}}$.  This is the correct order of magnitude to produce, given the uncertainties involved in estimating the magnitude of the acceleration and its possible variation with time, the observed velocity dependence with height of the stars in Per~OB1.

The proposed scenario provides a natural explanation for the relatively large height of HD~15137 above the galactic plane, and its motion.  \citet{boy05} suggest that both the massive stars HD~15137 and HD~14633 are runaway stars ejected from the open cluster NGC 654 in the Perseus spiral arm.  This suggestion, however, has a serious timescale problem, with inferred flight times for both stars that are longer than their ages \citep{boy05}.  If HD~15137 is a member of Per~OB1 as proposed in $\S 2.1$, then in our proposed picture it formed close to its present location and inherited its relatively large velocity away from the galactic plane from its swept-up parent molecular cloud.  Note that the O star lying at the next highest galactic latitude, HD~15137 (location as indicated in Fig.~2), has a comparably large proper motion away from the galactic plane.  This explanation, however, cannot be simply applied to HD~14633, which we also proposed in $\S 2.1$ to be a possible member of Per~OB1.  This star is located at a galactic latitude of $-18^{\circ}$, which is far beyond the presumed outer edge of the superbubble at a galactic latitude of about $-11^{\circ}$.

The triggered formation of stars in the cloud-shuffling model naturally predicts an age gradient from the center of the superbubble, such that stars at larger radii have younger ages.  If this model is applicable, the supergiant stars in Per~OB1 should therefore be closer to the $h$ and $\chi$~Per double cluster (the presumed center of the superbubble) than the O stars.  In Figure~9 we plot the spatial distributions of O stars (dotted lines) and supergiant stars (dashed lines) proposed to be members of Per~OB1 from \citet{hum78}.  Using the Kolmogorov-Smirnov test \citep{pres07}, we find a $65\%$ probability that these two populations have different distributions in galactic height, such that the younger O tend to be located higher above the galactic plane than the older supergiant stars (the visual impression given in Fig.~9).  Thus, there does seem to be a tendency, albeit only of marginally significance, for younger stars to be located at larger heights above the galactic plane.

\subsection{Recent Star Formation at High Galactic Latitudes}

In the cloud shuffling model as applied to Per~OB1, the relatively recent (within the last $\sim$1~Myr) formation of low- to intermediate-mass stars in remnant molecular clouds at unusually large heights above the galactic plane simply reflects the transport of molecular gas to such heights by the expanding superbubble.  The dust and \ion{H}{1} shells described in $\S3.1$ extend to galactic latitudes of about $-11^{\circ}$ (height $\sim$ 450~pc above the galactic plane), and so the remnant molecular clouds described in $\S 2.3.2$ lies in projection inside the superbubble.  These clouds may actually comprise swept-up molecular gas located at the periphery of the superbubble.  Alternatively, they may comprise the densest condensations in the giant molecular cloud that were not completely swept up by the expanding superbubble.

As mentioned in $\S2.3.2$, all the RMCs have head-tail morphologies with their heads pointing towards the galactic plane rather than in the direction of any particular O star.  These morphologies of these RMCs may therefore have been sculpted by the expanding superbubble rather than ionizing radiation from O stars.

\section{SUMMARY AND CONCLUSIONS}
The Per~OB1 association comprises the remarkable $h$ and $\chi$~Per double cluster (among the brightest and densest open clusters known), surrounded by luminous O/B and both early- and late-type supergiant stars.  Unlike virtually all other known OB associations in the Galaxy, there is no known giant molecular cloud in the vicinity of Per~OB1.

We have gathered together the available information on the stellar population of Per~OB1, primarily from the work by \citet{hum78} and \cite{gar92}.  For those luminous early- and late-type stars included in the Hipparcos astrometry mission, we have extracted measurements of their proper motions.  We found that:

\begin{enumerate}

\item these stars exhibit a bulk motion away from the galactic plane; and furthermore

\item the average velocity away from the galactic plane increases with height above the galactic plane

\end{enumerate}

We also searched for candidate intermediate-mass (HAeBe) and low-mass (Classical T~Tauri Star; CTTS) pre-main-sequence stars in Per~OB1.  We find a number of these candidate stars to be associated with four candidate remnant molecular clouds (RMCs) identified as dust clouds in {\it IRAS} 100-$\mu$m images and optical nebulae in the 2nd Digitized Sky Survey.  All these RMCs are located high (280--400~pc in projection) above the galactic plane.  From our spectroscopy of both candidate pre-main-sequence as well as main-sequence B--A stars apparently associated with these RMCs, we:

\begin{enumerate}

\item[3.] show that the distance moduli to the main-sequence B--A stars are consistent with the RMCs being located in Per~OB1

\item[4.] confirm that some of the candidate pre-main-sequence stars are indeed HAeBe stars and identify possible CTTSs, implying that there has been recent (within the last 1~Myr) star formation at galactic heights far above the locations of known luminous members in Per~OB1

\end{enumerate}

To explain the unusual properties of Per~OB1, we:

\begin{enumerate}

\item point out that the large shell of atomic hydrogen (\ion{H}{1}) gas (mass $\sim$$10^5 {\rm \ M_\sun}$) enclosing a slightly smaller shell of dust, studied by \citet{cap00} and which they propose to be created by a stellar-wind-blown superbubble from the $h$ and $\chi$~Per double cluster and OB stars in Per OB1, likely comprises (in part) the swept-up remains of the giant molecular cloud from which Per~OB1 formed

\item argue that if the observed motion of the luminous member stars is attributed to the dissolution of the $h$ and $\chi$~Per double cluster (at their outskirts), this can only explain the motion of the main-sequence B and late-type supergiant stars.  The main-sequence O stars, which are much younger than the double cluster, would not have had sufficient time to reach their present locations high above the galactic plane if they originated from the close vicinity of the double cluster

\item show that the observed motion of the luminous member stars is better explained in the cloud-shuffling model \citep{elm79,elm92}, in which the abovementioned superbubble pushes and accelerates the giant molecular cloud outwards and at the same time induces star formation at the dense cloud interface.  This would naturally result in those stars forming locally in the compressed molecular gas to have increasingly large velocities away from galactic plane with height

\item show that there is a weak trend for the younger O member stars to lie at higher galactic latitudes than the older late-type supergiant member stars, consistent with the simplest prediction of the cloud-shuffling but not cluster dissolution model

\item suggest that the recent star formation in the remnant molecular clouds that we found at very large heights above the galactic plane, far above the locations of the luminous member stars, likely occurred in the swept-up molecular gas at the periphery of the superbubble, or in the densest condensations that were not entirely swept away by the expanding superbubble

\end{enumerate}

\acknowledgments
H.-T. Lee thanks Wen-Ping Chen for sharing observation time, Wen-Shan Hsiao for helping spectral observations, Rue-Ron Hsu and Pin-Gao Gu for discussing, and the staff at the Beijing Astronomical Observatory for their assistance during observing runs.  We are grateful to the anonymous referee for constructive suggestions.  This research makes use of data products from the Two Micron All Sky Survey, which is a joint project of the University of Massachusetts and the Infrared Processing and Analysis Center/California Institute of Technology, funded by the National Aeronautics and Space Administration and the National Science Foundation.  This research made use of Montage, funded by the National Aeronautics and Space Administration's Earth Science Technology Office, Computation Technologies Project, under Cooperative Agreement Number NCC5-626 between NASA and the California Institute of Technology. Montage is maintained by the NASA/IPAC Infrared Science Archive.  This research made use of data products from the Digitized Sky Surveys produced at the Space Telescope Science Institute.

\clearpage

\begin{figure}
\includegraphics[angle=0,scale=0.8]{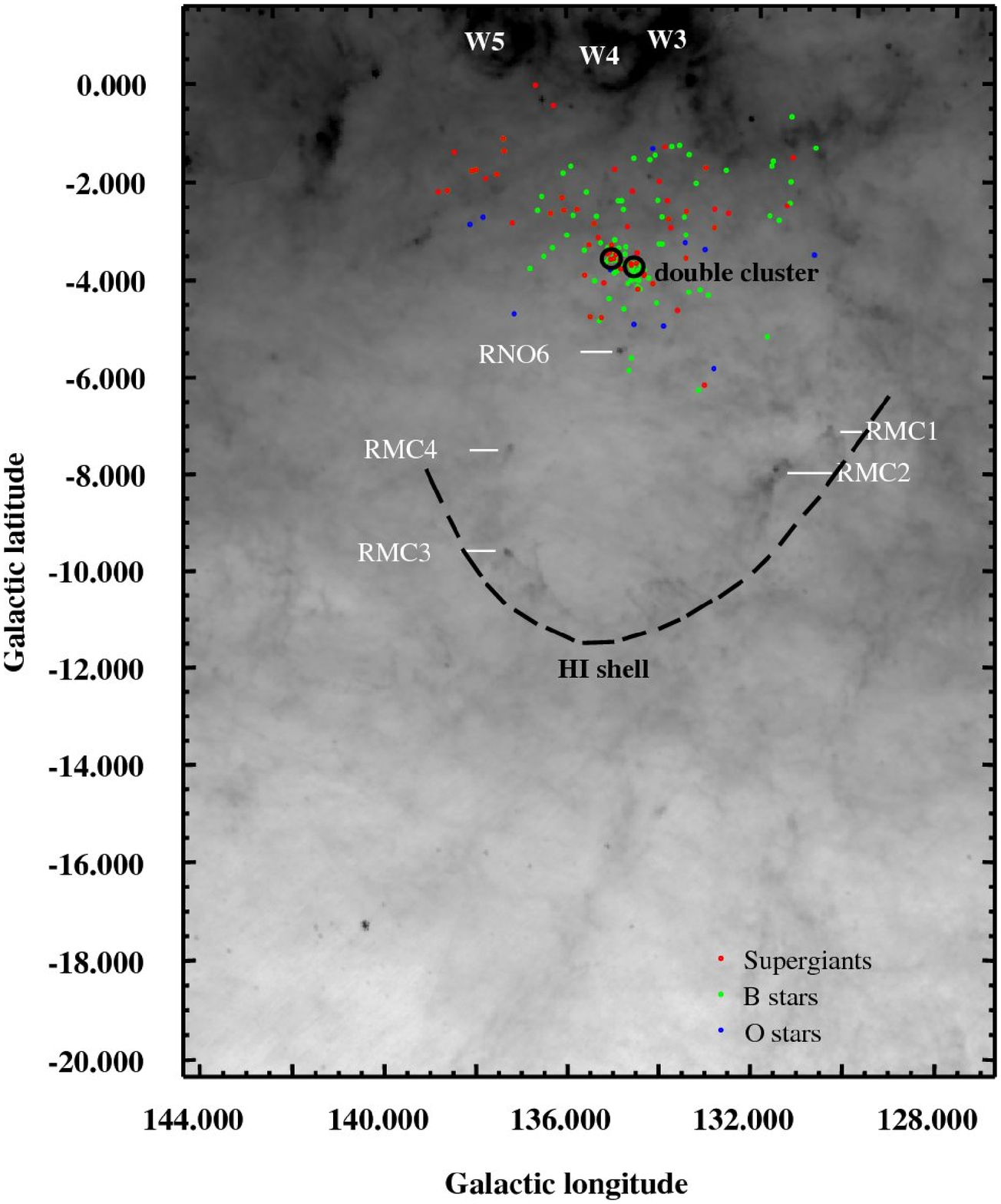}
\caption{Proposed luminous members of Per~OB1 from \cite{hum78} and \citet{gar92}, with different stellar types plotted in different colors.  The two black circles indicate the locations and extents of the $h$~Per (east) and $\chi$~Per (west) double cluster.  The locations of remnant molecular clouds (RMCs) identified in this work, as well as the remnant molecular cloud RNO6 are indicated.  The dashed line indicates the approximate geometry of the \ion{H}{1} shell found by \citet{cap00} and described in \S3.1.}
\end{figure}
\clearpage

\begin{figure}
\includegraphics[angle=0,scale=0.8]{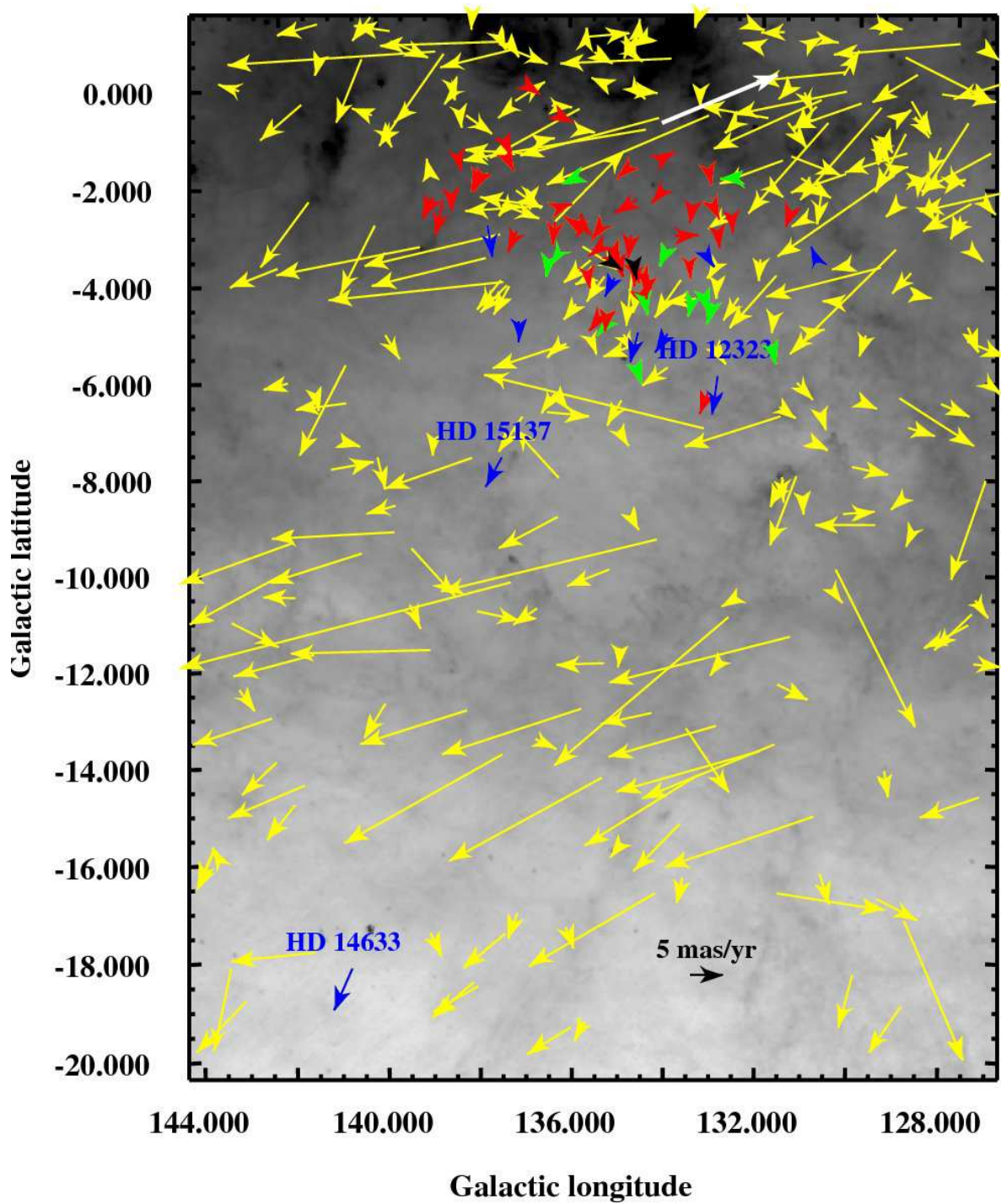}
\vspace{0.5mm}
\caption{Proper motions of the luminous members in Per~OB1 included in the {\it Hipparcos} catalog.  The blue, green, and red arrows are O, B, and supergiant stars respectively, and the two black arrows indicate the proper motions of the $h$ and $\chi$~Per double cluster.  Yellow arrows are for other stars in the {\it Hipparcos} catalog lying in the region shown.  The length of each arrow is proportional to the magnitude of its proper motion.  The grayscale is the $IRAS$ 100~$\micron$ image.}
\end{figure}
\clearpage

\begin{figure}
\includegraphics[angle=90,scale=0.6]{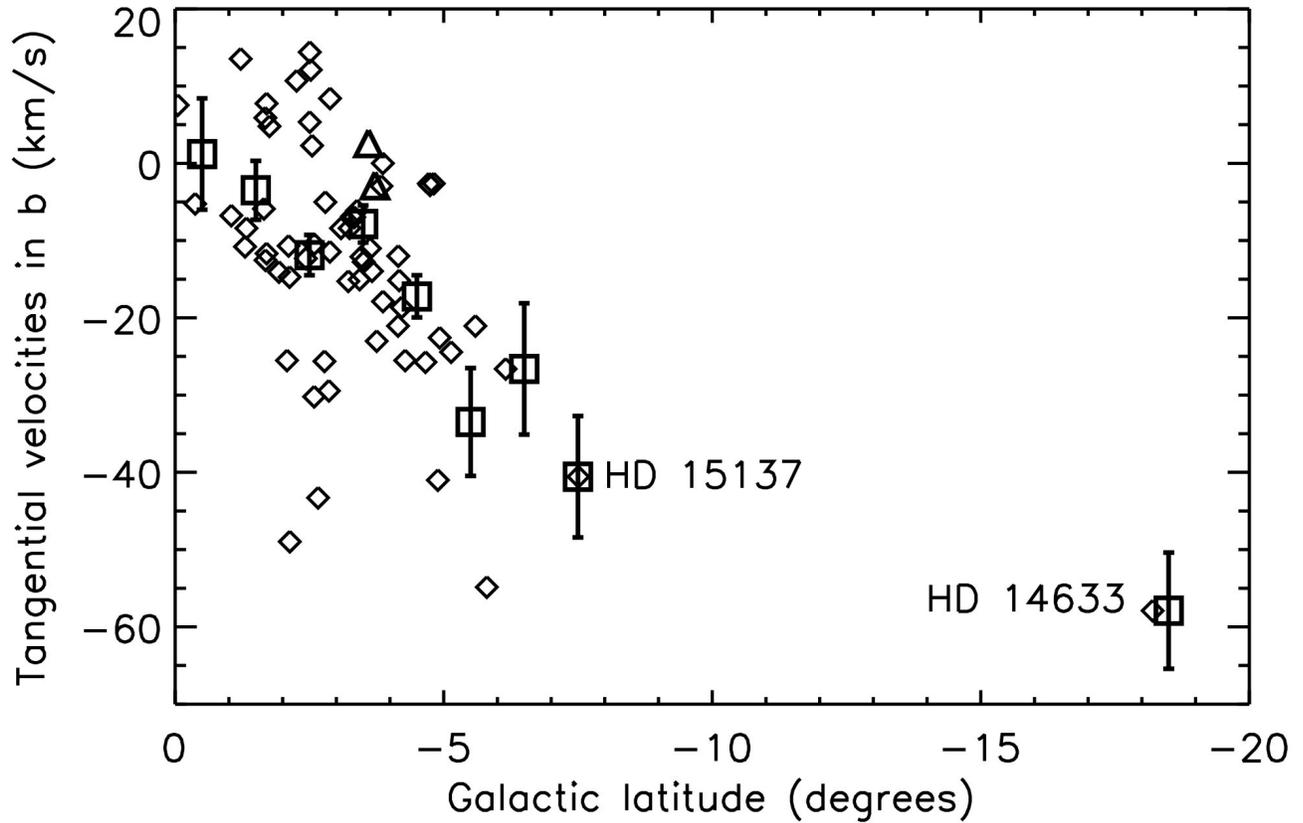}
\caption{Component of velocity along galactic latitude (tangential velocity) inferred from {\it Hipparcos} proper motions for all the proposed member stars (diamonds), as well as the $h$ and $\chi$~Per double cluster (triangles) by \citet{kha05}.  Also plotted are the tangential velocities for all proposed members in the {\it Hipparcos} averaged over latitude strips of $1^{\circ}$ (squares).  Note that all of the proper motions have been corrected for the solar motion.}
\end{figure}
\clearpage

\begin{figure}
\includegraphics[angle=90,scale=0.8]{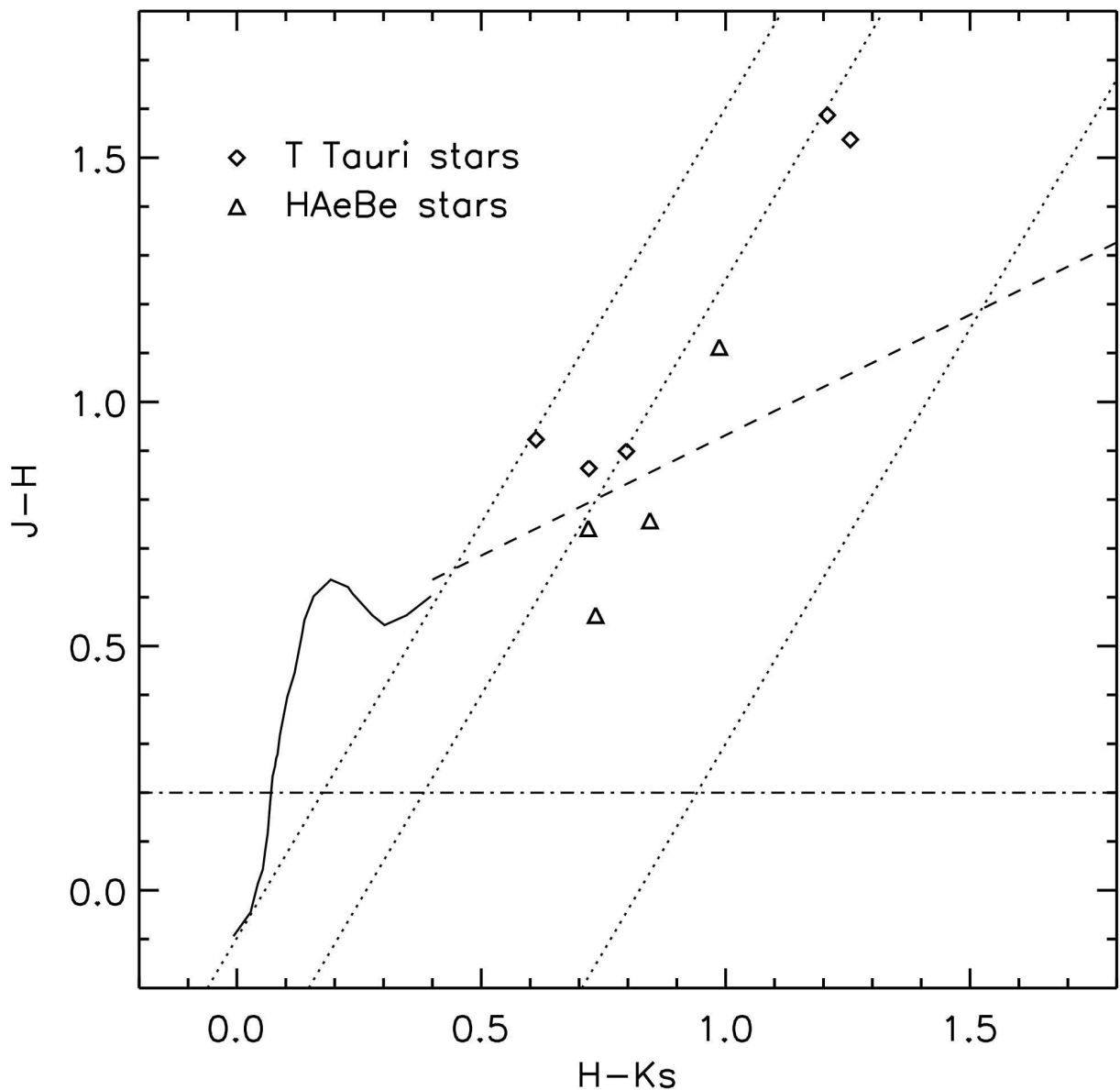}
\caption{Color-color diagram based on 2MASS measurements for our spectroscopically-confirmed pre-main-sequence stars.  Diamonds are CTTSs and triangles for HAeBe stars selected by 2MASS colors as listed in Table~4.   The solid line is the main sequence locus.  CTTS candidates lie between the two leftmost parallel dotted lines, and above the dereddened CTTS locus \citep{mey97} as indicated by the dashed line.  HAeBe star candidates lie between the two right-most parallel dotted lines and above the dash-dot horizontal line.}
\end{figure}
\clearpage

\begin{figure}
\includegraphics[angle=0,scale=0.9]{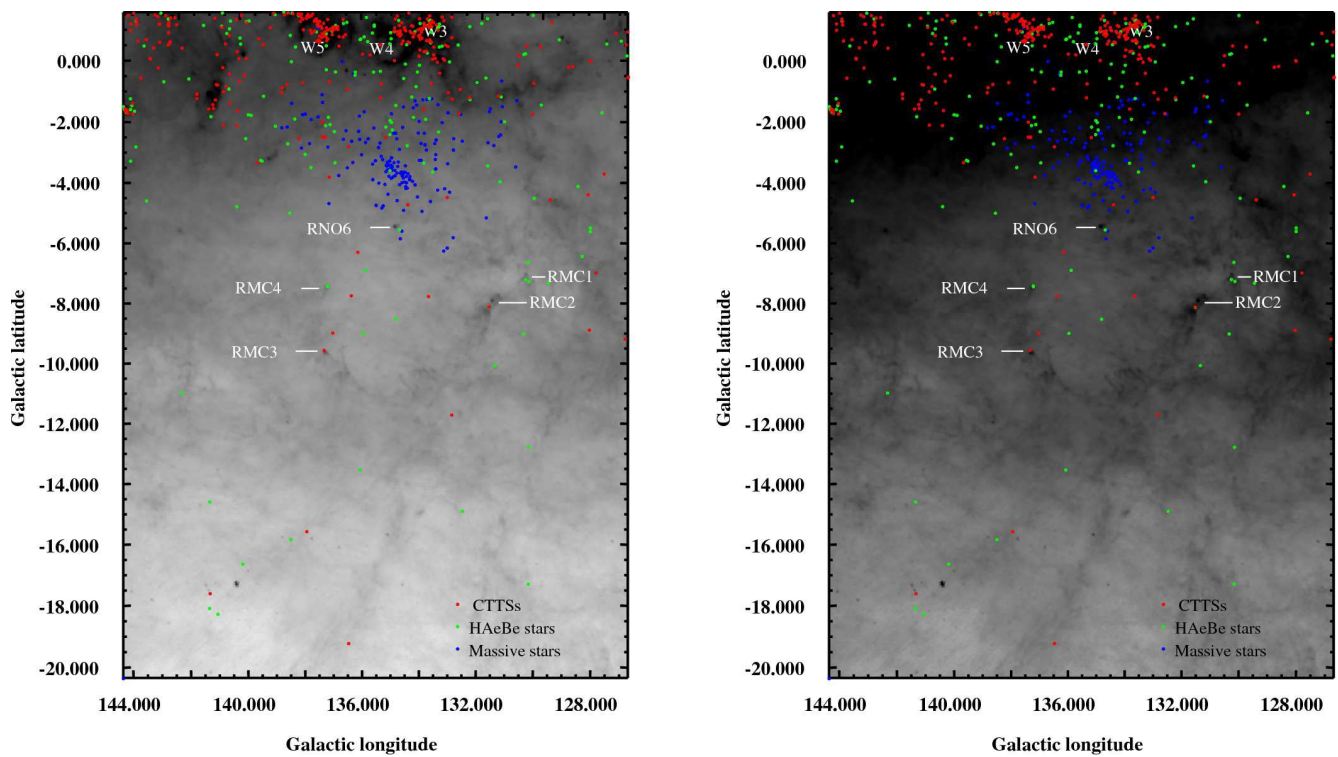}
\caption{Candidate CTTS (red) and HAeBe (green) stars selected in this work, along with proposed luminous members (blue) of Per~OB1, plotted on the $IRAS$ 100~$\micron$ image using two different stretches.  The left panel emphasizes regions with the strongest extinction, and the right panel regions with lower extinctions to more clearly show the RMCs identified in this work.  Note that only those candidate CTTS and HAeBe stars lying close to or against the RMCs were selected for follow-up spectroscopy.}
\end{figure}
\clearpage

\begin{figure}
\includegraphics[angle=0,scale=0.4]{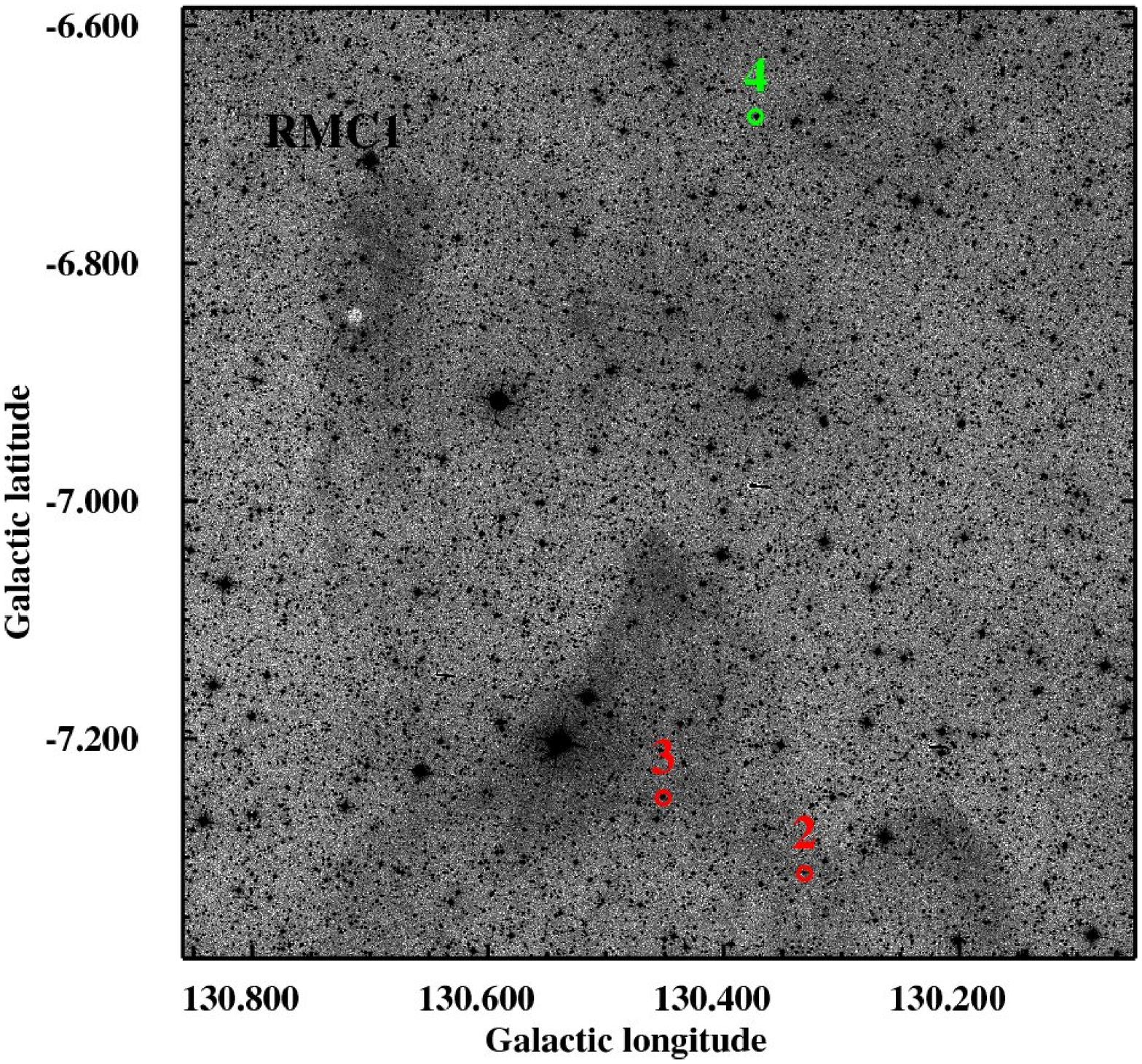}
\includegraphics[angle=0,scale=0.4]{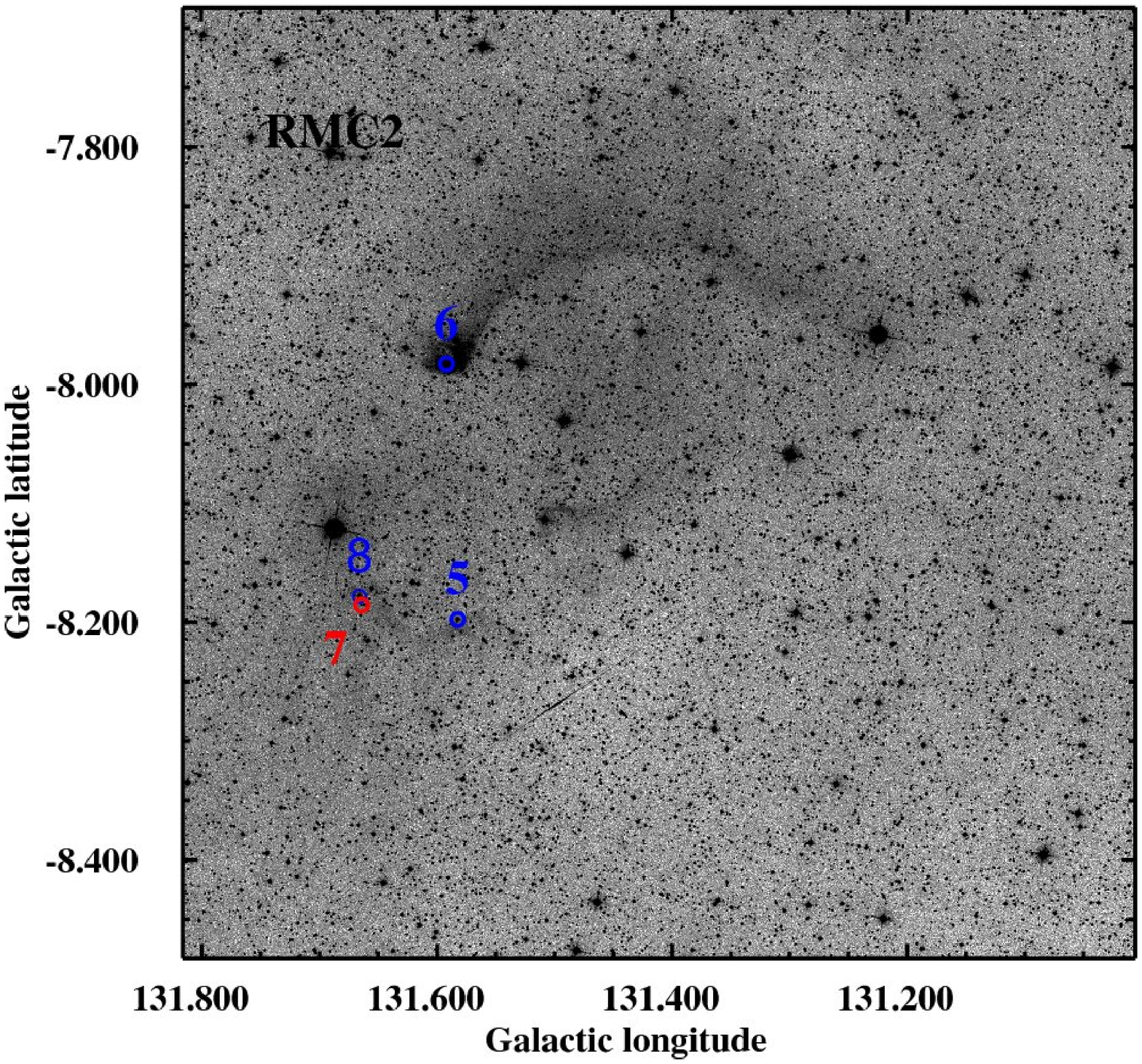}
\includegraphics[angle=0,scale=0.4]{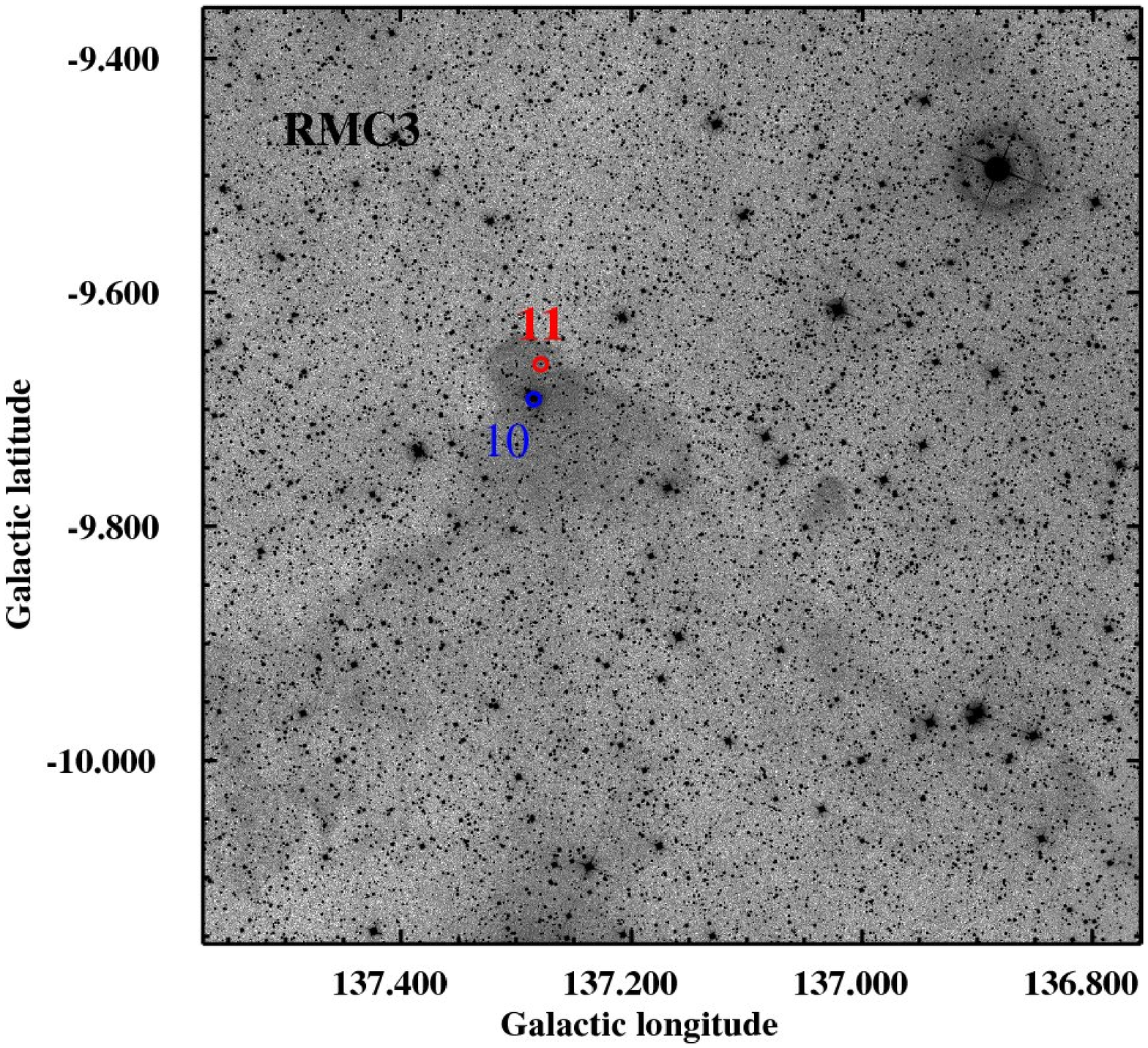}
\includegraphics[angle=0,scale=0.4]{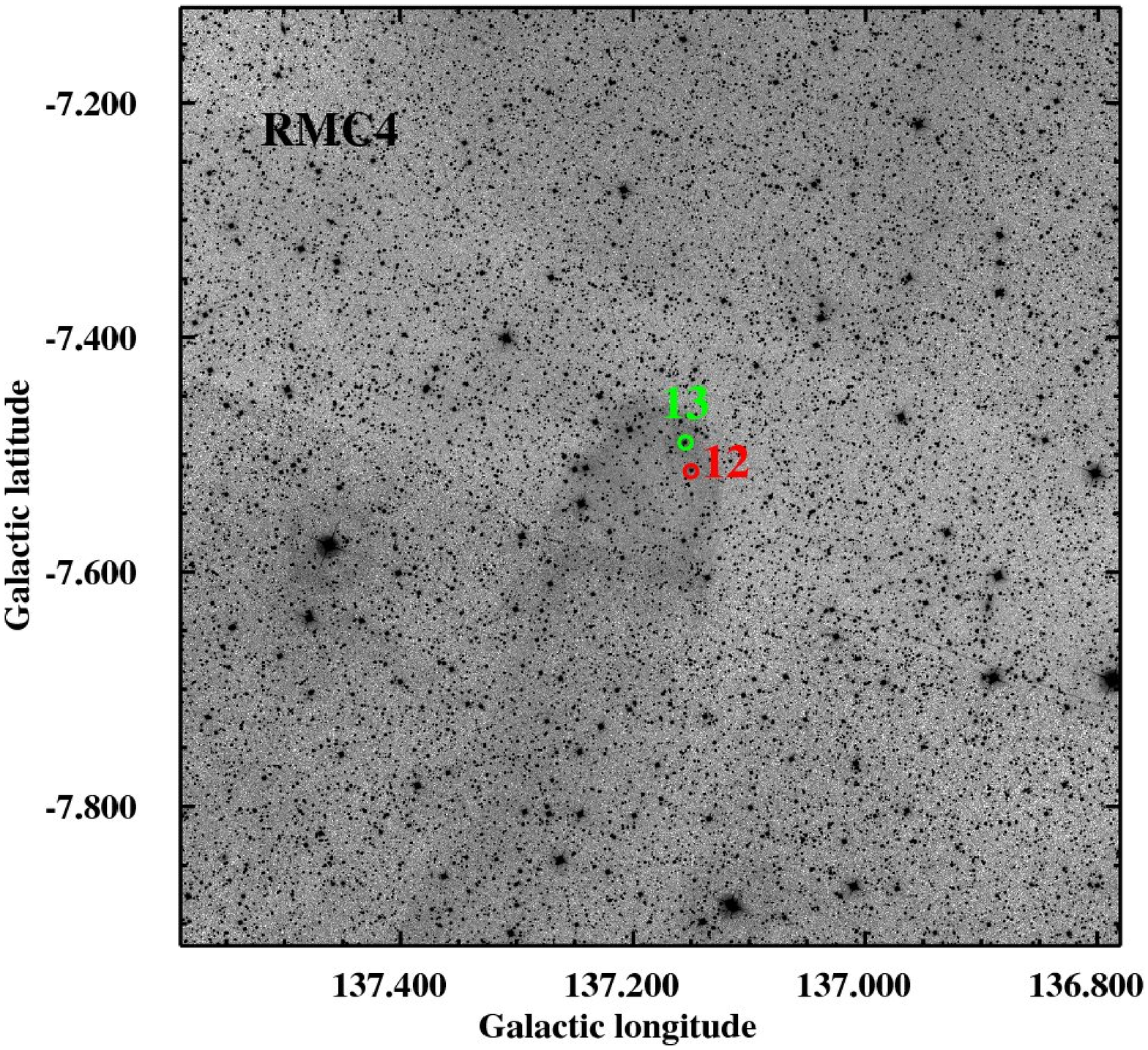}
\includegraphics[angle=0,scale=0.4]{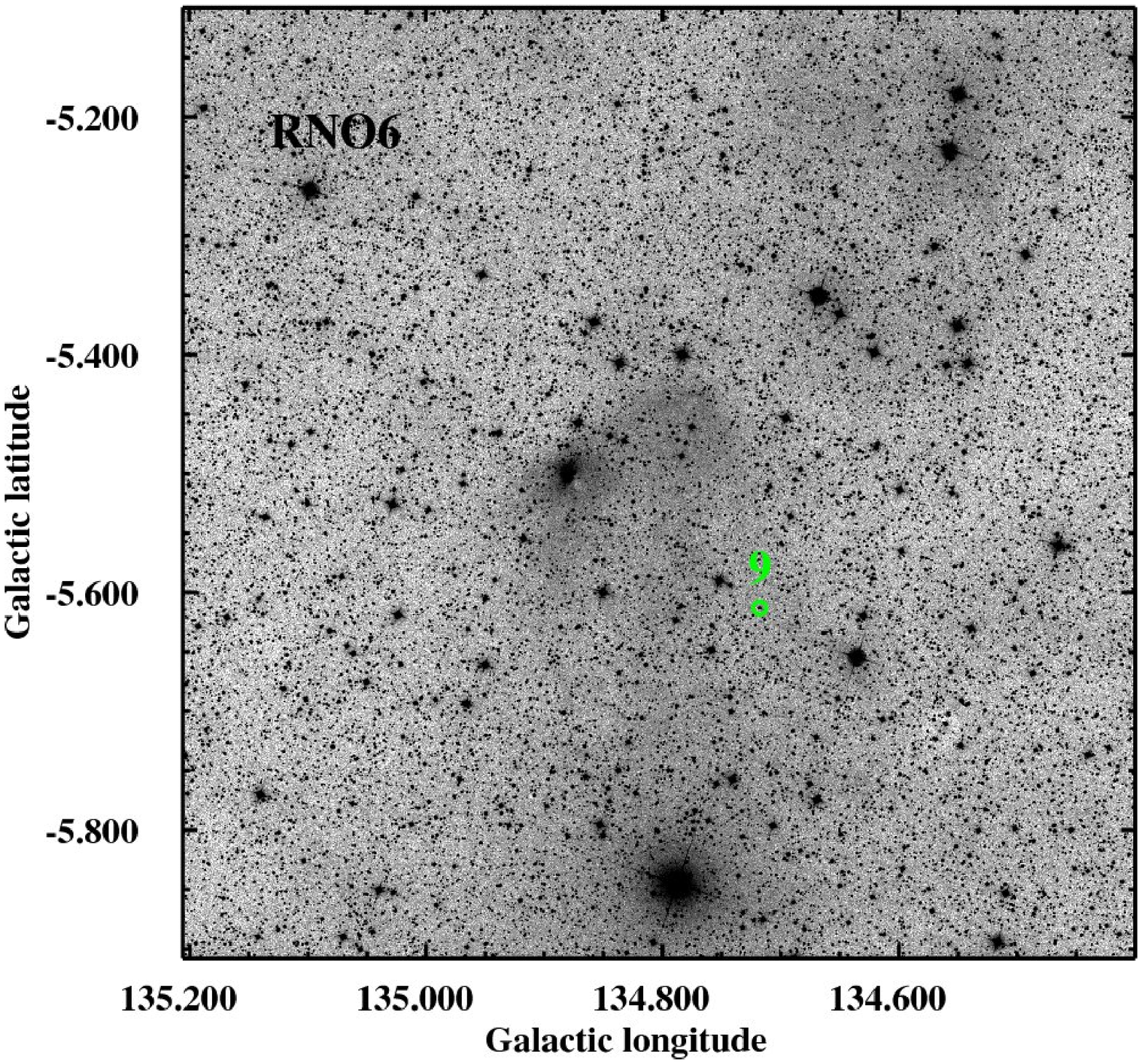}
\caption{Spectroscopically identified CTTSs, HAeBe stars, and B or A stars. Closeup of individual candidate RMCs as seen in R-band images from the 2nd Digitized Sky Survey.  The CTTSs, HAeBe stars, and B or A stars are labeled in red, green, and blue, respectively.}
\end{figure}
\clearpage

\begin{figure}
\includegraphics[angle=0,scale=0.8]{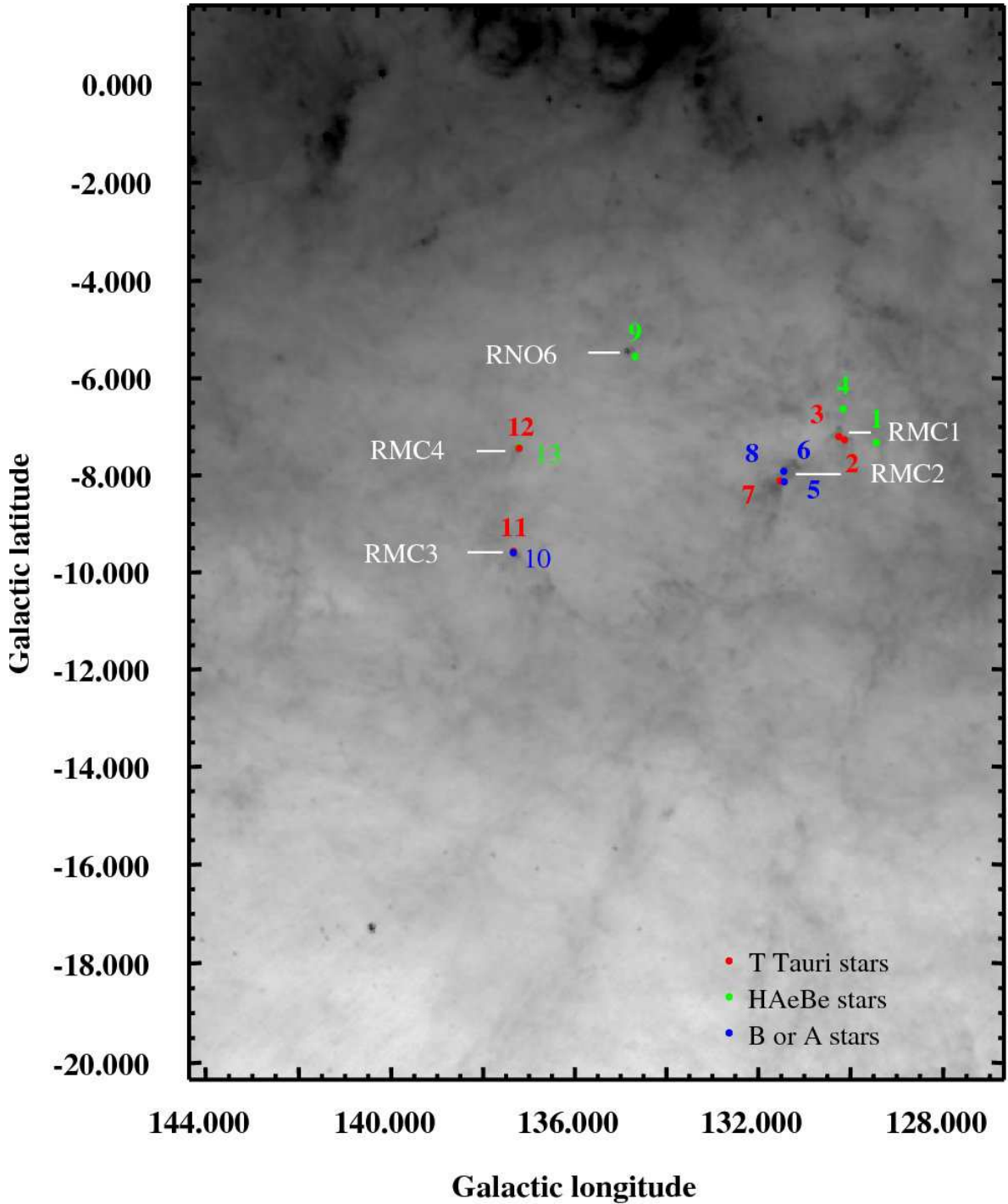}
\vspace{3mm}
\caption{Our spectroscopically identified CTTS (red) and HAeBe (green) stars plotted on the $IRAS$ 100~$\micron$ image.  Locations of the RMCs identified in this work, as well as RNO6, are indicated.  Also indicated are the locations of main-sequence B or A stars (blue) used to estimate distances to the RMCs.  The stars are numbered according to their identification numbers in Table~4.}
\end{figure}
\clearpage

\begin{figure}
\includegraphics[angle=90,scale=0.6]{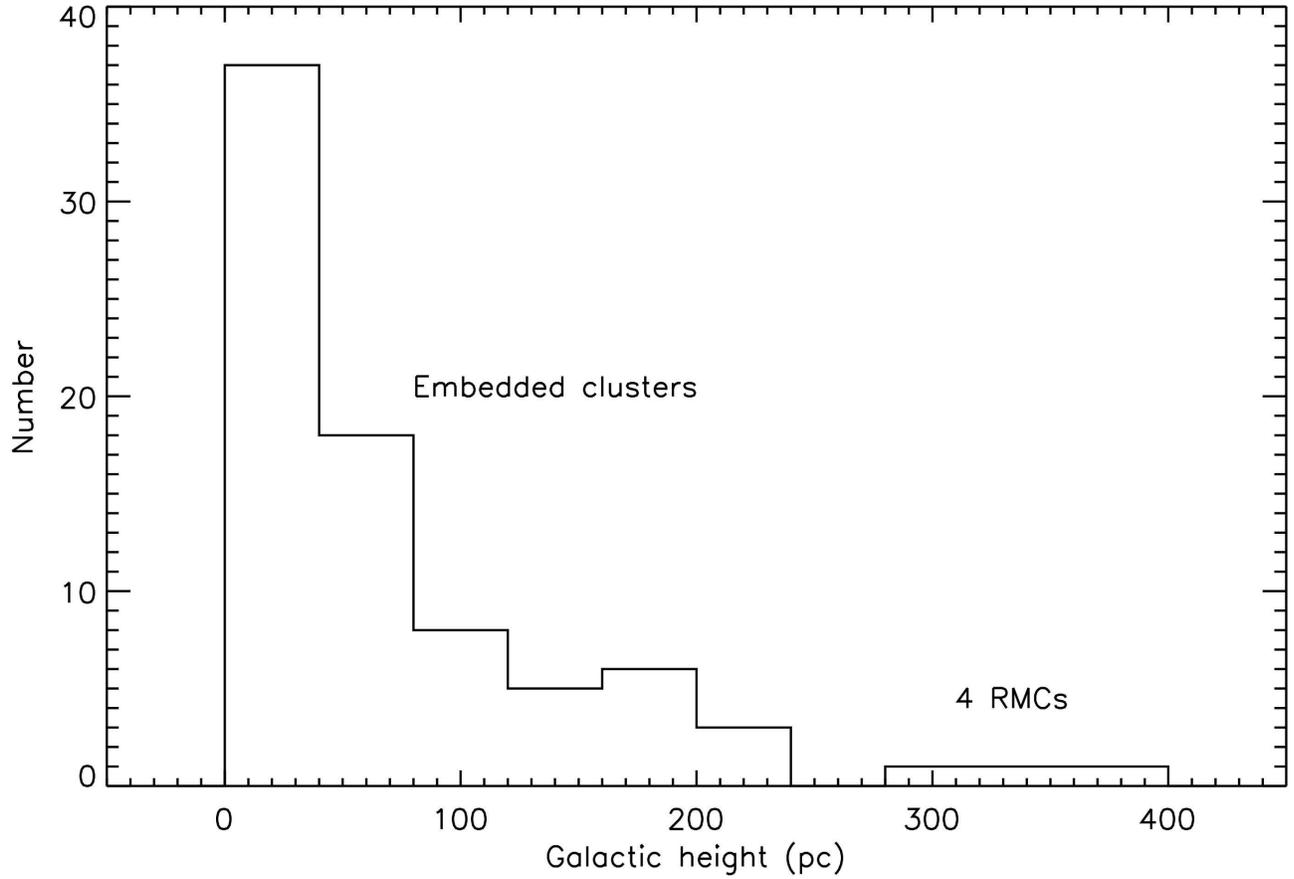}
\caption{Number of embedded clusters as a function of galactic height as compiled by \citet{lad03}, along with the RMCs identified in this work.  Note that all the previously known embedded cluster are confined within 240~pc of the galactic plane, whereas the four newly identified RMCs are located beyond 280~pc.}
\end{figure}
\clearpage

\begin{figure}
\includegraphics[angle=90,scale=0.6]{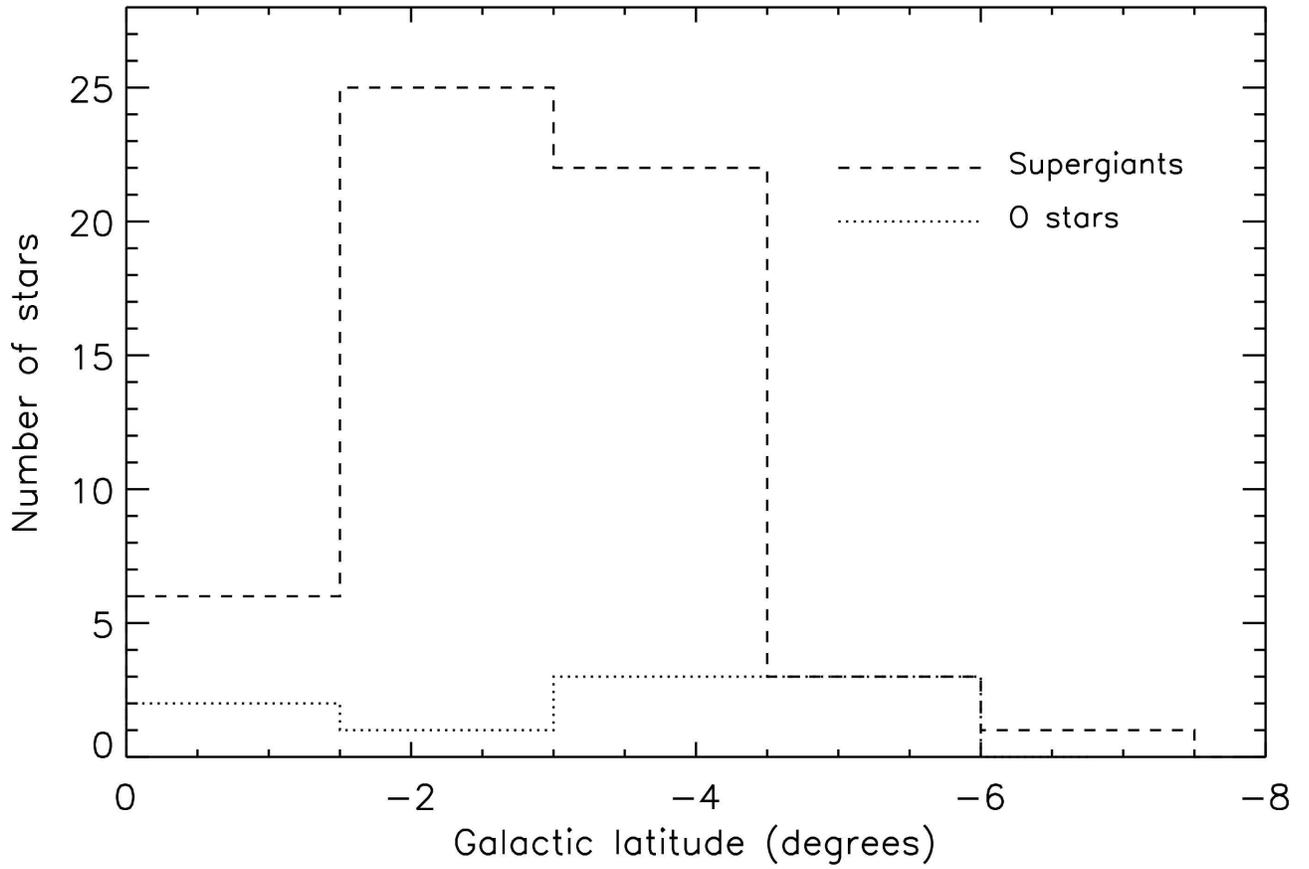}
\caption{Number of O (dotted line) and supergiant (dashed line) stars in \citet{hum78} plotted as a function of galactic latitude.  There is a weak trend for the younger O stars to lie, as a whole, at higher galactic latitudes than the older supergiant stars.}
\end{figure}
\clearpage


\begin{deluxetable}{rrrrrrrrll}
\tablecaption{Members of Per OB1 with Hipparcos Proper Motions}
\rotate
\tablewidth{0pt}
\tablehead{\colhead{$l$} & \colhead{$b$} & \colhead{Hip} & \colhead{pm$l^{a}$}  & \colhead{pm$b^{a}$}  & \colhead{e\_pm$l^{a}$} & \colhead{e\_pm$b^{a}$} & \colhead{star} & \colhead{Sp Type} & \colhead{Ref.}  \\
           \colhead{deg} & \colhead{deg} &               & \colhead{mas yr$^{-1}$} & \colhead{mas yr$^{-1}$} & \colhead{mas yr$^{-1}$}  & \colhead{mas yr$^{-1}$} &                &                   &              }
\startdata

130.900 &  -3.441 &  8725 &  1.05 &  2.49 & 1.38 & 1.13 &  HD 236894 &  O8V             & a, b \\
131.479 &  -2.433 &  9211 &  0.65 & -1.88 & 1.03 & 0.87 &  HD 236915 &  M2Iab           & b    \\
131.867 &  -5.138 &  9017 & -0.85 & -3.01 & 1.46 & 1.33 &  BD+55 441 &  B1Vpe           & a    \\
132.691 &  -2.588 &  9886 & -0.12 & -1.71 & 1.11 & 1.05 &  HD 236947 &  M2Iab-Ib        & b    \\
132.742 &  -1.700 & 10078 &  1.40 & -0.03 & 1.49 & 1.10 &  HD 13036  &  B0.5III         & a, b \\
132.975 &  -5.802 &  9523 &  0.77 & -5.80 & 1.00 & 1.02 &  HD 12323  &  ON9V            & a, b \\
132.977 &  -2.504 & 10060 & -0.13 & -0.26 & 1.00 & 0.98 &  HD 13022  &  O9.5Ia          & a, b \\
132.981 &  -2.880 &  9990 & -0.30 & -1.80 & 0.58 & 0.61 &  HD 12953  &  A1Ia            & a, b \\
133.105 &  -4.282 &  9826 &  0.72 & -3.10 & 1.12 & 0.98 &  HD 12727  &  B2III           & a    \\
133.173 &  -6.153 &  9582 &  1.25 & -3.21 & 0.71 & 0.78 &  HD 12401  &  M4Ib            & b    \\
133.168 &  -1.650 & 10323 & -0.30 & -1.28 & 1.49 & 1.10 &  HD 13402  &  B0.5Ib          & a, b \\
133.174 &  -3.334 & 10024 & -0.87 & -1.39 & 1.16 & 0.99 &  HD 12993  &  O6.5V           & a, b \\
133.270 &  -4.170 &  9935 & -0.51 & -2.15 & 0.94 & 0.87 &  HD 12856  &  B0IIIpe         & a, b \\
133.505 &  -4.217 & 10055 &  0.72 & -2.47 & 1.09 & 0.98 &  HD 13051  &  B1IVe           & a    \\
133.565 &  -2.549 & 10379 &  0.09 & -0.54 & 0.65 & 0.65 &  HD 13476  &  A3Iab           & a, b \\
133.571 &  -3.509 & 10227 & -0.14 & -1.92 & 0.66 & 0.61 &  HD 13267  &  B5Ia            & a, b \\
133.888 &  -2.882 & 10489 & -3.66 &  0.02 & 1.16 & 1.11 &  HD 13658  &  M1Iab           & b    \\
134.013 &  -1.219 & 10904 & -0.90 &  0.50 & 1.69 & 1.43 &  HD 14242  &  M2Iab           & b    \\
134.030 &  -4.924 & 10228 &  1.90 & -2.83 & 0.92 & 0.74 &  HD 13268  &  O7:             & a, b \\
134.064 &  -3.223 & 10527 &  0.90 & -2.15 & 1.03 & 0.90 &  HD 13716  &  B0.5III         & a, b \\
134.130 &  -1.931 & 10829 &  1.78 & -2.03 & 1.36 & 1.08 &  HD 14142  &  M2Iab           & b    \\
134.445 &  -3.868 & 10624 &  0.23 & -2.39 & 0.77 & 0.69 &  HD 13841  &  B2Ib            & a, b \\
134.448 &  -3.840 & 10633 & -0.60 & -1.02 & 0.54 & 0.52 &  HD 13854  &  B1Iab           & a, b \\
134.525 &  -4.149 & 10615 & -0.51 & -2.68 & 0.86 & 0.81 &  HD 13831  &  B0IIIe          & a, b \\
134.571 &  -4.154 & 10641 & -0.85 & -1.85 & 0.68 & 0.66 &  HD 13866  &  B2Ib            & a, b \\
134.649 &  -4.891 & 10541 &  1.16 & -4.52 & 0.83 & 0.76 &  HD 13745  &  O9.7II((N))     & a, b \\
134.685 &  -2.129 & 11093 &  3.21 & -2.09 & 1.98 & 1.71 &  HD 14528  &  M4Ie            & b    \\
134.705 &  -5.589 & 10443 & -0.90 & -2.69 & 0.99 & 0.92 &  HD 13621  &  B0.5III-IV((N)) & a    \\
134.705 &  -3.665 & 10805 &  0.24 & -2.03 & 0.68 & 0.59 &  HD 14134  &  B3Ia            & a, b \\
134.714 &  -3.626 & 10816 & -0.64 & -1.76 & 0.66 & 0.59 &  HD 14143  &  B2Ia            & a, b \\
134.792 &  -2.861 & 10995 &  0.56 & -3.45 & 0.84 & 0.77 &  HD 14404  &  M2Iab           & b    \\
135.033 &  -3.433 & 11020 & -0.50 & -2.13 & 0.61 & 0.57 &  HD 14433  &  A1Ia            & a, b \\
135.058 &  -1.678 & 11394 &  0.21 & -0.20 & 1.33 & 0.97 &  HD 14947  &  O5If+           & a, b \\
135.120 &  -3.247 & 11115 & -0.84 & -1.52 & 0.72 & 0.63 &  HD 14542  &  B8Ia            & a, b \\
135.143 &  -3.752 & 11018 &  1.46 & -2.86 & 0.99 & 0.84 &  HD 14434  &  O6.5:           & a, b \\
135.155 &  -3.385 & 11098 & -1.34 & -1.31 & 0.75 & 0.66 &  HD 14535  &  A2Iab           & a, b \\
135.334 &  -4.743 & 10926 &  0.10 & -1.02 & 0.79 & 0.63 &  HD 14322  &  B8Ib            & a, b \\
135.381 &  -4.815 & 10935 &  0.90 & -1.00 & 1.05 & 0.83 &  HD 14331  &  B0III           & a, b \\
135.400 &  -3.088 & 11284 &  2.05 & -1.52 & 1.32 & 1.36 &  HD 14826  &  M2Iab           & b    \\
135.488 &  -2.795 & 11391 &  0.94 & -1.21 & 0.62 & 0.59 &  HD 14956  &  B2Ia            & a, b \\
135.573 &  -4.724 & 11060 &  0.76 & -0.99 & 0.66 & 0.54 &  HD 14489  &  A2Iab           & a, b \\
135.683 &  -3.866 & 11279 & -0.13 & -0.75 & 0.51 & 0.50 &  HD 14818  &  B2Ia            & a, b \\
135.835 &  -2.506 & 11625 &  0.07 &  0.58 & 0.68 & 0.63 &  HD 15316  &  A2Ia            & a, b \\
136.119 &  -2.524 & 11769 &  0.55 &  0.37 & 0.73 & 0.67 &  HD 15497  &  B6Ia            & a, b \\
136.128 &  -1.754 & 11953 &  0.73 & -0.30 & 1.21 & 0.98 &  HD 15752  &  B0IIIN          & a, b \\
136.152 &  -2.260 & 11841 & -0.92 &  0.24 & 1.01 & 0.90 &  HD 15620  &  B8Iab           & a, b \\
136.323 &  -0.371 & 12416 & -2.23 & -1.21 & 1.23 & 1.13 &  BD+58 501 &  M2Iab           & b    \\
136.351 &  -3.293 & 11722 &  1.10 & -1.38 & 1.22 & 1.14 &  HD 15450  &  B1IIIe          & a    \\
136.385 &  -2.587 & 11898 &  0.33 & -3.51 & 1.88 & 1.75 &  HD 15690  &  B2Iab           & a, b \\
136.535 &  -3.468 & 11782 &  0.34 & -1.86 & 1.37 & 1.30 &  HD 15548  &  B1V             & a    \\
136.695 &  -0.046 & 12750 & -0.06 & -0.04 & 0.74 & 0.67 &  HD 16778  &  A2Ia            & b    \\
137.159 &  -4.660 & 11850 &  0.05 & -3.11 & 1.11 & 1.05 &  HD 15642  &  O9.5III:N       & a, b \\
137.185 &  -2.779 & 12302 &  1.33 & -3.09 & 0.86 & 0.86 &  HD 236979 &  M1Iab           & b    \\
137.351 &  -1.300 & 12769 & -0.84 & -1.73 & 1.50 & 1.20 &  HD 16808  &  B0.5Ib          & a, b \\
137.367 &  -1.044 & 12840 & -0.58 & -1.36 & 1.27 & 1.09 &  HD 236995 &  A0Ia            & a, b \\
137.532 &  -7.507 & 11473 &  2.53 & -4.48 & 0.67 & 0.72 &  HD 15137  &  O9.5III(N)      & c    \\
137.796 &  -2.660 & 12676 & -0.65 & -4.71 & 1.04 & 1.14 &  HD 16691  &  O5f             & a    \\
137.932 &  -1.681 & 12972 &  1.01 & -1.89 & 0.94 & 0.93 &  HD 17088  &  B9Ia            & a, b \\
138.022 &  -1.701 & 13022 &  0.69 & -1.81 & 1.10 & 1.13 &  HD 17145  &  B8Ia            & a, b \\
138.383 &  -1.326 & 13290 & -0.36 & -1.51 & 1.61 & 1.46 &  HD 237010 &  M2Iab           & b    \\
138.537 &  -2.110 & 13178 & -0.03 & -1.72 & 0.63 & 0.69 &  HD 17378  &  A5Ia            & a, b \\
138.718 &  -2.133 & 13262 &  1.21 & -5.23 & 1.01 & 1.03 &  HD 237008 &  M3Iab           & b    \\
138.930 &  -2.081 & 13380 &  1.49 & -3.08 & 1.44 & 1.54 &  HD 17638  &  WC6             & c    \\
140.863 & -18.126 & 11099 &  3.13 & -6.05 & 0.87 & 0.69 &  HD 14633  &  ON8V            & c    \\
\enddata
\tablenotetext{a}{The proper motions and errors in the Hipparcos catalog are converted from the equatorial coordinates to the galactic coordinates.}
\label{table:members}
References--(a) \citet{gar92}; (b) \citet{hum78}; (c) this paper
\end{deluxetable}


\begin{deluxetable}{rrrrrr}
\tablecaption{}
\tablewidth{0pt}
\tablehead{\colhead{$b$} & \colhead{average pm in $l^{a}$} & \colhead{average pm in $b^{a}$} & \colhead{e\_pml} & \colhead{e\_pmb} & \colhead{number of stars} \\ 
\colhead{degree} & \colhead{mas/yr} & \colhead{mas/yr} & \colhead{mas/yr} & \colhead{mas/yr}}
\startdata
   0 -- $-1$   &  $-1.02$    &     $0.11$     &   0.72 & 0.66  &  2  \\
$-1$ -- $-2$   &    0.44     &    $-0.32$     &   0.42 & 0.35  & 11  \\
$-2$ -- $-3$   &    0.46     &    $-1.09$     &   0.26 & 0.24  & 18  \\
$-3$ -- $-4$   &    0.41     &    $-0.72$     &   0.24 & 0.22  & 16  \\
$-4$ -- $-5$   &    0.67     &    $-1.58$     &   0.28 & 0.25  & 11  \\
$-5$ -- $-6$   &    0.06     &    $-3.07$     &   0.68 & 0.64  &  3  \\
$-6$ -- $-7$   &    1.64     &    $-2.44$     &   0.71 & 0.78  &  1  \\
$-7$ -- $-8$   &    2.58     &    $-3.72$     &   0.67 & 0.72  &  1  \\
......... \\
$-18$ -- $-19$ &    2.93     &    $-5.31$     &   0.87 & 0.69  &  1  \\
\enddata
\tablenotetext{a}{proper motions have been corrected for the solar motion}
\label{table:pm}
\end{deluxetable}


\begin{deluxetable}{lcccc}
\tablecaption{Remnant Molecular Clouds}
\tablewidth{0pt}
\tablehead{\colhead{RMC} & \colhead{$l$} & \colhead{$b$} & \colhead{Galactic Height} & \colhead{LBN}\\
                                                    & \colhead{degree} & \colhead{degree} & \colhead{pc} & }
\startdata
RMC1    & 130.45 & -7.12 & 287 & 638 \\
RMC2    & 131.43 & -7.98 & 322 & 640 \\
RMC3    & 137.29 & -9.70 & 393 & 666 \\
RMC4    & 137.20 & -7.53 & 304 & --  \\

\enddata
\label{table:rmc}
\end{deluxetable}

\begin{deluxetable}{lcccl}
\tablecaption{Spectral Observations}
\tabletypesize{\footnotesize}
\tablewidth{0pt}
\tablehead{\colhead{Star} & \colhead{2MASS} & \colhead{H$\alpha$ Emission Line$^{a}$} & \colhead{Type} & \colhead{Remarks}\\}

\startdata
1  & J01380790+5453463 & H$^{b}$    & HAeBe (A0) &  around RMC1                                  \\ 
2  & J01424061+5449135 & H($-$52.5) & CTTS       &   in RMC1                          \\ 
3  & J01433441+5451323 & H($-$6.1)  & CTTS       &   in RMC1                          \\ 
4  & J01435138+5526098 & H$^{b}$    & HAeBe (B3) &   in RMC1                          \\ 
5  & J01494099+5341341 &            & B (B7V)    &   with a reflection nebula in RMC2 \\ 
6  & J01500456+5354007 &            & B (B3V)    &   with a reflection nebula in RMC2 \\ 
7  & J01501404+5341107 & H($-$89.1) & CTTS       &   in RMC2                          \\ 
8  & J01501546+5341350 &            & A (A2V)    &   with a reflection nebula in RMC2 \\ 
9  & J02151126+5519500 & H$^{b}$    & HAeBe (B8) &   around RNO 6                     \\ 
10 & J02220484+5038130 &            & B (B8V)    &   with a reflection nebula in RMC3 \\ 
11 & J02220662+5040013 & H($-$2.2)  & WTTS       &   in RMC3                          \\ 
12 & J02261522+5243131 & H($-$16.1) & CTTS       &   with a reflection nebula in RMC4                         \\ 
13 & J02262049+5244288 & H$^{b}$    & HAeBe (B6V)&   with a reflection nebula in RMC4 \\ 
\enddata
\tablenotetext{a}{The numbers following H are the equivalent widths of H$\alpha$ of T Tauri Stars.}
\tablenotetext{b}{H$\alpha$ line is in absorption but filled in by emission.}
\label{table:spectra}
\end{deluxetable}
\clearpage

\end{document}